\documentclass[floatfix,aps,pra,twocolumn,a4paper,superscriptaddress,nofootinbib,balancelastpage,longbibliography]{revtex4-1}
\usepackage[english]{babel}
\usepackage{amsopn,amsthm,dsfont,latexsym}

\usepackage{mathtools}
\mathtoolsset{showonlyrefs=true}
\usepackage{tikz}
\usetikzlibrary{positioning}

\usepackage[caption=false]{subfig}

\begin{document}
\title{Causality -- Complexity -- Consistency:\\
Can Space-Time Be Based on Logic and Computation?}
\author{\"Amin Baumeler}\author{Stefan Wolf}
\affiliation{Faculty of Informatics, Universit\`{a} della Svizzera
  italiana, Via G. Buffi 13, 6900 Lugano, Switzerland}
 \affiliation{Facolt\`a indipendente di Gandria, Lunga scala, 6978 Gandria, Switzerland}

\begin{abstract}
	\noindent
The difficulty of explaining non-local correlations in a fixed causal
structure sheds new light on the old debate on whether space and time 
are to be seen as fundamental. Refraining from assuming space-time
as given  {\em a priori\/} has a number of consequences. First, the usual definitions of 
{\em randomness\/} depend on a causal structure and turn meaningless.
So motivated, we propose an {\em intrinsic},  physically motivated measure for the
randomness of
a string of bits: its length minus its normalized work value, a quantity we  closely relate 
to its {\em Kolmogorov complexity\/} (the length of the
  shortest 
program making a universal Turing machine output this string).
We test this alternative concept of randomness
for the example of non-local correlations, and we end up
with a
reasoning
that leads to similar conclusions as in, but is conceptually more direct than,
the probabilistic view since only the outcomes of measurements that can 
{\em actually all  be carried out together\/} are put into relation to
each 
other. In the same context-free spirit, we connect the logical
reversibility of an evolution to the second law of 
thermodynamics and the arrow of time. Refining this, we end up
with a speculation on the emergence of a space-time structure on 
bit strings in terms of data-compressibility relations. Finally, we
show that  logical consistency, by which we replace the 
abandoned  causality, it strictly weaker a constraint than the
latter
in the multi-party case.
\end{abstract}

\maketitle

\section{Randomness Without Causality}\label{eins}

What is
{\em causality\/}?~| The notion has been defined in different ways and turned out
to be  highly problematic,  both in Physics and
Philosophy.
 This observation is not new, as is nicely shown by {\em Bertrand Russell\/}'s
quote~\cite{russell} from more than a century ago:
\\ 

\noindent
{\it 
``The law of causality [\ldots]\ is a relic of a bygone age, surviving,
like the monarchy, only because it is erroneously supposed to do no
harm.''
}
\\

Indeed, a number of attempts have been made to abandon causality and
replace global by only local assumptions (see, {\em
  e.g.},~\cite{ocb}). A particular motivation is  given by the
difficulty of explaining quantum non-local correlation according to
{\em Reichenbach's principle}~\cite{Reichenbach}. The latter states
that in a given (space-time)
causal 
structure, correlations
stem
from {\em a common cause\/} (in the common past) or a {\em direct influence\/} from 
one of the events  to the other.  In the case of
violations of Bell's inequalities, a number of results indicate that
explanations
through some mechanism as suggested by 
Reichenbach's principle either fail to
explain 
the correlations~\cite{bell} or are unsatisfactory since they  require
infinite speed~\cite{beforebefore},~\cite{coretti},~\cite{jdb},~\cite{tomy} or precision~\cite{ws}. All of 
this may serve as a motivation for dropping the assumption 
of a global causal structure in the first place.

Closely related to causality is the notion of {\em
  randomness\/}: In~\cite{colren}, a piece of information is called 
{\em freely random\/} if it is statistically independent from all 
other pieces of information except the ones in its future light cone.
Clearly, when the assumption of an initially given causal structure 
is dropped, such a definition is not possible any longer. One may
choose
to consider  freely random pieces of {\em information \/} as
being more fundamental than a space-time structure~| in fact, the
latter can then be seen as emerging from the former: If a piece 
of information is free, then any piece correlated to it is in its
causal future.\footnote{This change of perspective 
reflects the debate, three centuries ago, between {\em Newton\/} and {\em Leibniz\/} on the nature 
of space and time, in particular on  as how fundamental this causal
structure is to be considered.}
But how can we {\em define\/} the randomness of an object {\em purely
  intrinsically\/} and independently of any context?

For further motivation, note that Colbeck and Renner's definition of randomness~\cite{colren} is
consistent with full determinism: 
A random variable with trivial distribution is independent of every
other
(even itself). How can we exclude this and
additionally ask for the possibility in principle of a counterfactual outcome,
{\em i.e.}, 
that the random variable $X$ {\em could have taken a value different from
the one it actually took\/}? 
Intuitively, this is a necessary condition for {\em freeness}.
The question whether the universe (or a {\em closed\/} system) 
starting from a given state $A$ always ends up in {\em the same\/} state
$B$ seems to be meaningless: Even if rewinding were possible, and two 
runs could be performed, the outcomes $B_1$ and $B_2$ that must be 
compared never exist in the same reality since  rewinding
erases the result of the rewound run~\cite{renner}: ``$B_1=B_2$?'' is 
{\em not a question which cannot be answered in principle, but that
  cannot even be formulated precisely}. 
In summary, defining freeness of a choice or a  random event,
understood as the {\em actual possibility of two (or more) well-distinguishable options}, seems hard even when a causal structure {\em is\/} in place.\footnote{In this
  context and as a reply to~\cite{gisin}, we feel that the notion of a
  {\em choice between different possible futures
  by an act of free
  will\/} put forward there is not only hard to formalize but also not
much more innocent  than Everettian 
relative states~\cite{everett}~| after all, the latter {\em are\/}  real  (within their respective
branches of the wave function).
We have become familiar with the ease of handling probabilities and cease to realize how delicate they are ontologically.}

We look for an {\em intrinsic\/} definition of randomness that
takes into account only the ``factuality,'' {\em i.e.}, the state of
the closed system in question. 
Clearly, such a definition is hard to imagine for a 
single bit, but it {\em can\/} be defined in a natural way for (long) strings
of 
bits, namely 
its length minus the work value (normalized through dividing by~$kT$)
of a physical representation of the 
string with respect to some extraction device;
we relate this quantity to the string's ``best compression.''

We test the alternative view of randomness for physical 
meaning. More specifically, we find it to be 
functional in the context of {\em non-local
correlations\/}: A reasoning
yielding a similar mechanism as in 
the probabilistic regime is realized which has the conceptual
advantage not to require  relating the outcomes of 
measurements that cannot all actually be carried out.
That mechanism is: Random inputs to a non-local system plus 
no-signaling guarantee random outputs. 

In the second half of this text,  we consider consequences of 
abandoning (space-time) causality as being fundamental. 
In a nutshell, we put {\em logical reversibility\/} to the center of our 
attention here. We argue that if a
computation on a Turing machine is logically reversible, then a
``second law'' emerges: The complexity of the tape's content cannot 
decrease in time. 
This law holds without failure probability, in contrast to the 
``usual'' second law, and implies the latter.
 In the same 
spirit, we propose to define causal relations between  physical
points, modeled by bit strings, as given by the fact that ``the past is entirely contained in
the future,'' {\em i.e.}, nothing is forgotten.\footnote{It has been
  argued that quantum theory violates the causal law due to random
  outcomes of measurements. Grete Hermann~\cite{gh} argued that 
the law of causality does not require the past to determine the
future,
but {\it vice versa}. This is in accordance with our view of logical
reversibility:
There can be information growth, but there can be no information loss.}
In this view, we also study the relationship between full causality
(which we aim at dropping)
and mere 
logical consistency (that we never wish to abandon) in the complexity
view: They are different from each other as soon as more than two
parties are involved.

\section{Preliminaries}

Let ${\cal U}$ be a fixed universal Turing machine (TM).\footnote{The introduced asymptotic notions
are independent of this choice.}
For a finite
or infinite string $s$,  the {\em Kolmogorov
  complexity\/}~\cite{kol},~\cite{text} \mbox{$K(s)=K_{\cal U}(s)$} is the length
of the shortest program for~${\cal U}$ such that the machine outputs~$s$. Note that $K(s)$ can be infinite if~$s$ is.

Let $a=(a_1,a_2,\ldots)$
be an infinite string. Then
\[
a_{[n]}:=(a_1,\ldots,a_n,0,\ldots)\ .
\]
We study the asymptotic behavior of $K(a_{[n]})\, :\, {\bf
  N}\rightarrow {\bf N}$. For this function, we simply write~$K(a)$,
similarly $K(a\, |\, b)$ for $K(a_{[n]}\, |\, b_{[n]})$,  the latter being the length of 
the shortest program outputting~$a_{[n]}$ upon input~$b_{[n]}$. 
We write
\[
K(a)\approx n\ :\Longleftrightarrow\ \lim_{n\rightarrow\infty}\left(
\frac{K(a_{[n]})}{n}\right) =1\ .
\]
We call a string $a$ with this property {\em  incompressible}. 
We also use $K(a_{[n]})=\Theta(n)$, as well as
\[
K(a)\approx 0 :\Longleftrightarrow \lim_{n\rightarrow\infty}\left(
 \frac{K(a_{[n]})}{n}\right) = 0 \Longleftrightarrow K(a_{[n]})=o(n).
\]
Note that {\em computable\/} strings $a$ satisfy $K(a)\approx 0$, and
that incompressibility is, in this sense, the extreme case of uncomputability.

Generally, for  functions $f(n)$ and $g(n)\not\approx0$, we write 
$f\approx g$ if $f/g\rightarrow 1$.
{\em Independence of $a$ and $b$\/} is  then\footnote{This is inspired by Cilibrasi and Vit\'{a}nyi~\cite{cbc}, where (joint) Kolmogorov complexity~--- 
or, in practice, any efficient compression method~--- is used to
define 
a distance measure on sets of  bit strings (such as literary texts of
genetic information of living beings). The resulting structure in that
case is a distance measure, and ultimately a clustering as a binary tree.}
\[
K(a\, |\, b)\approx K(a)
\]
or, equivalently, 
\[
K(a,b)\approx K(a)+K(b)\ .
\]
If we introduce
\[
I_K(x;y):=K(x)-K(x\, |\, y)\approx K(y)-K(y\, |\, x)\ ,
\]
independence of $a$ and $b$ is $I_K(a,b)\approx 0$.

In the same spirit, we can define {\em conditional independence\/}: We
say that
{\em $a$ and $b$ are independent given $c$\/} if 
\[
K(a,b\, |\, c)\approx K(a\, |\, c)+K(b\, |\, c)
\]
or, equivalently, 
\[
K(a\, |\, b,c)\approx K(a\, |\, c)\ ,
\]
or
\[
I_K(a;b\, |\, c):=K(a\, |\, c)-K(a\, |\, b,c)\approx 0\ .
\]

\section{Complexity as Randomness 1: Work Extraction}\label{fuel}

\subsection{The Converse of Landauer's Principle}

In our search for an {\em intrinsic\/} notion of randomness~| independent of 
probabilities or the existence of alternatives~| expressed through the
properties of the object in question, we must realize, first of all, 
that such a notion is impossible for single bits, since neither of the 
two possible values, $0$~nor~$1$, is in any way more an argument for the 
``randomness'' of that bit than not. The situation, however, changes
for {\em long\/}
strings of bits: No one would call the one-million-bit string
$000\cdots 0$ {\em random\/} (even though, of course, it is not impossible
that this string originates from a random process such as a million
consecutive tosses of a fair coin). In the spirit of Rolf
Landauer's~\cite{landau98}
famous slogan ``information is
physical,'' we may want to test our intuition physically: If the $N$ bits
in a  string encode the position, being in the left ($0$) as opposed the
right ($1$)  half 
of some container, of the molecules of a gas, then the $0$-string
means that the gas is all concentrated in one half and, hence, allows
for extracting
work from the environmental
heat; the amount is $NkT\ln 2$ if $k$ is Boltzmann's constant and~$T$ is the temperature of the
environment. This fact has also been called the {\em converse of 
Landauer's principle}. 
Note that any other system which can be transformed by a reversible
process into that maximally asymmetric gas has the same work value; an
example
 is a physical representation of 
 the first $N$ bits of the binary expansion of $\pi$ of the same
length~| although this string may look much more ``random'' at first
sight.  
This reversible process is, according to the {\em
  Church-Turing thesis},\footnote{The Church-Turing thesis, first
  formulated by Kleene~\cite{CT}, states that any physically possible
  process can be simulated by a universal Turing machine.}
imagined to be carried out by a Turing
machine in such a way that every step is {\em logically reversible\/}
(such as, {\em e.g.}, a Toffoli gate) and  
can be {\em un\/}computed by the same device;  the process is then 
also possible in principle in a
{\em thermodynamically\/} reversible way: No heat is dissipated~\cite{ballist}. 
It is  clear that most $N$-bit strings cannot have any 
work value  provided there is no {\em perpetuum mobile of the second
kind}. 

For a given string $S$, its {\em length minus the work value of a
  physical representation\/} (divided by $kT$) may be regarded as an 
intrinsic measure for the {\em randomness\/} of $S$. 
We address the question what in general the fuel value is of
(a physical representation  of)  $S$. 
Since (the {\em reversible\/} extraction of) the 
string $0^N$ from $S$ is equivalent to (the gain of) free energy of
$NkT\ln 2$, we have a first  answer:
{\em 
Work  extraction is  data compression}.

\subsection{Free Energy and Data Compression}

\begin{figure}[h]
	\centering
	\includegraphics[scale=1]{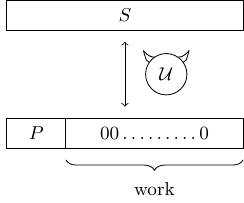}
	\caption{Bennett's argument.}
	\label{fig:1}
\end{figure}

\noindent
{\it State of the art.}
Bennett~\cite{bennetttoc} 
claimed the fuel value of a   string $S$ 
to be {\em its length minus~$K(S)$\/}:
\[
W(S)=({\rm len}(S)-K(S))kT\ln2\, .
\]
Bennett's argument is that (the physical representation of) $S$ can be
| logically, hence, thermodynamically~\cite{ballist}~| reversibly transformed into the string
$
P||000\cdots 0
$,
where $P$ is the shortest program for ${\cal U}$ generating $S$
and  the length of the generated $\mbox{$0$-string}$ is 
${\rm  len}(S)-K(S)$ (see Figure~\ref{fig:1}). 

It was already pointed out by Zurek~\cite{zurek} that 
whereas it is true that the {\em reverse
direction\/} exists and is {\em computable\/} by a universal Turing machine, 
its {\em forward direction}, {\em i.e.},~$P$ 
from~$S$, is {\em not}. This means that the demon that can
 carry out the  work-extraction computation on~$S$ from scratch
 does   not  physically exist if the Church-Turing hypothesis is true. We will see,
 however, that
Bennett's value  is  an {\em upper bound\/} on the
fuel
value of $S$.

Dahlsten {\em et al.}~\cite{dahlsten} follow
Szil\'{a}rd~\cite{szilard29} in putting the {\em knowledge\/}
of the demon extracting the work to the center of their attention.
More precisely, they claim  
\[
W(S)=({\rm len}(S)-D(S))kT\ln2\, ,
\]
where the ``defect'' $D(S)$ is bounded from above and below by a 
 smooth R\'enyi entropy of the
distribution of~$S$ from the demon's viewpoint,
modeling her ignorance.
They do not consider the algorithmic aspects of the demon's
actions  extracting the free energy, but  the effect 
of the demon's {\em a priori knowledge on $S$}. If~we model the  demon
 as an algorithmic apparatus, then we should
specify the {\em form\/} of
that knowledge explicitly: Vanishing conditional
entropy means that~$S$ is {\em uniquely determined\/} from the demon's viewpoint. 
Does this  mean that the demon possesses a {\em copy\/} of~$S$, or the 
{\em ability\/} to produce such a copy, or pieces of {\em information\/}
that uniquely determine~$S$? This question sits at the origin of the  gap
between the two described groups of results; it is maximal when
the demon fully  ``knows''~$S$  which, however, still has maximal
complexity even given her internal state (an example see below). 
In~this case, the first result claims $W(S)$ to
be $0$, whereas $W(S)\approx {\rm len}(S)$ according to the
second. The gap vanishes if
  ``knowing $S$'' 
is understood in a {\em constructive\/}~| as opposed to entropic~|
 sense, meaning that ``the demon
possesses or can produce
a copy of~$S$ represented in her internal state:'' If that copy is
included in Bennett's reasoning, then his
 result reads
\begin{align}
\frac{W(S,S)}{kT}&\approx {\rm len}(S,S)-K(S,S)\\
&\approx 2\, {\rm len}(S)-K(S)\\
&\approx {\rm len}(S)
\,.
\end{align}
In this case, knowledge has immediate work value. 

\

\begin{figure}[h]
	\centering
	\includegraphics[scale=1]{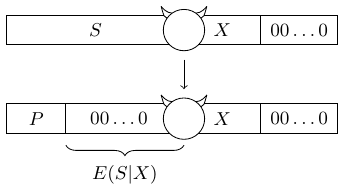}
	\caption{The model.}
	\label{fig:3}
\end{figure}

\noindent
{\it The model.}
We assume the {\em demon\/} to be 
a {\em universal Turing machine\/ ${\cal U}$\/} the memory tape of
which is  
sufficiently long for the tasks and inputs in question, but {\em
  finite}.
The tape initially contains~$S$, the string the fuel value of which is
to be determined, $X$, a~finite string modeling the  demon's {\em knowledge 
about~$S$}, and $0$'s for the  rest of the tape. After  the
 extraction computation, the tape contains, at the bit positions 
initially holding $S$, a (shorter) string $P$ plus 
$
0^{{\rm len}(S)-{\rm len}(P)}
$,
whereas 
the rest of the tape
is (again) the same  as before work extraction. 
The demon's operations are
{\em logically\/} reversible and can, hence, be carried out
{\em thermodynamically\/} reversibly~\cite{ballist}. 
Logical reversibility in our model is the 
ability of the same demon to carry out the backward computation step
by step,
{\em i.e.}, from $P||X$ to~$S||X$.\footnote{Note that this is the
  natural way of defining logical reversibility in our setting with a
  {\em fixed\/} input and output but {\em no sets nor bijective maps\/} between them.} 
We denote by~$E(S|X)$  the {\em maximal amount of $0$-bits 
extractable logically reversibly from $S$ given the knowledge~$X$},
{\em i.e.},
\[
E(S|X):={\rm len}(S)-{\rm len}(P)
\]
if $P$'s length is minimal
(see Figure~\ref{fig:3}).

\

According to the above, the work value of any physical representation
of $S$ for a demon knowing~$X$~is 
\[
W(S|X)= E(S|X)kT\ln 2\ .
\]

\noindent
{\it Lower bound on the fuel value.}
Let $C$ be a computable  function
\[
C\, :\, \{0,1\}^*\times \{0,1\}^* \longrightarrow \{0,1\}^*
\]
such that 
\[
(A,B)\mapsto (C(A,B),B)
\]
is injective. We call $C$ a {\em data-compression algorithm with
  helper}.
Then we have 
\[
E(S|X)\geq {\rm len}(S)-{\rm len}(C(S,X))\, .
\]

This can be seen as follows. First, note that the function
\[
A||B\ \mapsto\ C(A,B)||0^{{\rm len}(A)-{\rm len}(C(A,B))}||B
\]
is computable and bijective.
Given the two (possibly irreversible) circuits computing the compression
and its inverse, one can obtain a {\em reversible\/} circuit
realizing the function and
where no further input or output bits are involved. 
This can be achieved by first
implementing all logical operations with Toffoli gates and uncomputing  all
junk~\cite{bennettTM} in both of the circuits. The resulting two circuits have now
both still the property that the input is part of the output. 
As a second step, we can simply
combine the two, where the first circuit's first output becomes the 
second's second input, and {\it vice versa}. Roughly speaking,
the first circuit computes the compression and the second reversibly
uncomputes the raw data. The combined circuit has
only the compressed data (plus the~0's) as output, on the bit
positions carrying the input previously. 
(The depth of this circuit is
roughly 
the sum of the depths of the two irreversible
circuits 
for the compression and for the decompression, respectively.) 
We assume 
that circuit to be
hard-wired in the demon's head.
A~typical example for a  compression algorithm that can be used is
Ziv-Lempel~\cite{zl}.

\

\noindent
{\it Upper bound on the fuel value.}
We have the following upper bound on $E(S|X)$:
\[
E(S|X)\leq {\rm len}(S)-K_{\cal U}(S|X)\, .
\]
The reason is that  the demon is only  able to carry out the
computation in question (logically, hence, thermodynamically)
reversibly 
{\em if she is able to carry
  out 
the reverse computation  as well}.  Therefore, the string $P$ must
be at least as long as the shortest program for ${\cal U}$ generating~$S$ if $X$ is given.

Although the same is not true in general, this upper bound is
{\em tight\/} if $K_{\cal U}(S|X)= 0$.
The latter means that~$X$ itself is a program for generating
an additional
copy of~$S$. The demon can then bit-wisely XOR this new
 copy 
to the original  $S$ on the tape, hereby producing $0^{{\rm
    len}(S)}$
{\em reversibly\/} to replace the original~$S$ (at the same time preserving the new
one, as  reversibility demands). When Bennett's ``uncomputing
trick'' is used~|  allowing 
 for
making 
any computation by a Turing machine logically reversible~\cite{bennettTM}~|,
 then
a history string $H$ is written to the tape during the computation of
$S$ from~$X$ such that after the XORing, the demon can,
going back step by step,
{\em uncompute\/} the generated copy of $S$ and end up in the tape's 
original state~| except that the original~$S$ is now replaced by~$0^{{\rm
    len}(S)}$: This
results in a maximal fuel value matching the  (in this
case trivial) upper bound. Note that this  harmonizes
with~\cite{dahlsten} if
vanishing conditional entropy is so established. 

\

\begin{figure}[h]
	\centering
	\subfloat[\label{fig:2a}]{
		\includegraphics[scale=1]{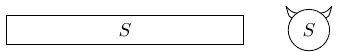}
	}
	\\
	\subfloat[\label{fig:2b}]{
		\includegraphics[scale=1]{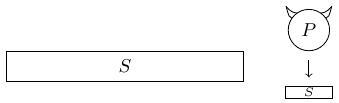}
	}
	\\
	\subfloat[\label{fig:2c}]{
		\includegraphics[scale=1]{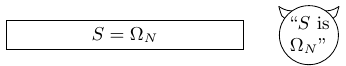}
	}
	\caption{Knowing~$S$.}
	\label{fig:2}
\end{figure}

\noindent
{\it Discussion.}
We contrast our bounds with the  entropy-based results
of~\cite{dahlsten}: According to the latter, a demon {\em 
having complete knowledge of $S$\/}
 is able to extract maximal work: $E(S)\approx
{\rm len}(S)$. 
{\em What does ``knowing $S$'' mean\/} (see Figure~\ref{fig:2})?
We have seen that
the results are in accordance with ours if the demon's 
{\em knowledge\/} consists of (a)~a {\em copy\/} of $S$, or at least of (b)~the
{\em ability to algorithmically 
reconstruct $S$}, based on a known program $P$, as 
discussed above. It is, however, possible (c)~that the demon's 
knowledge is of different nature, merely {\em determining $S$ uniquely
without providing the ability to  build $S$}. For instance,
let the demon's knowledge about $S$ be:
``$S$ equals the first $N$ bits~$\Omega_N$ of the binary expansion of~$\Omega$.'' 
Here, $\Omega$ is the so-called {\em halting probability\/}~\cite{chaitin} 
of a fixed
universal Turing machine ({\em e.g.}, the demon ${\cal U}$ itself).  Although
there 
is a  {\em short description\/} of $S$ in this case, 
and~$S$ is thus 
uniquely determined in an entropic sense, there is no  {\em set
  of instructions  shorter than $S$ enabling 
  the demon to
generate~$S$\/}~| which would be required for  work extraction
from~$S$ 
according to our upper bound. In short, this
gap 
reflects the one between the {\em ``unique-description complexity''\/}\footnote{A diagonal argument, called {\em Berry paradox}, 
 shows that the 
notion of ``description complexity'' cannot be defined generally for
all strings.} and
the {\em Kolmogorov complexity}.

\section{Complexity as Randomness 2: Non-Locality}\label{nl}

\subsection{Non-Locality from Counterfactual Definiteness}

\begin{figure*}
\centering
\subfloat[\label{bild1}]{
\includegraphics[scale=1]{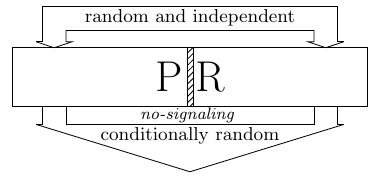}}
\qquad
\subfloat[\label{bild2}]{
\includegraphics[scale=1]{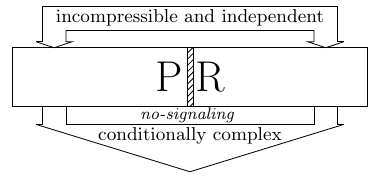}}
\caption{The traditional (a) vs.\ the new (b) view: Non-locality {\em
    \`a la\/} Popescu/Rohrlich (PR) plus
  no-signaling leads to the output inheriting {\em randomness\/} (a) or
  {\em complexity\/}~(b), respectively, from the input.}
\label{ganz}
\end{figure*}

Non-local correlations~\cite{bell} are a fascinating feature of quantum
theory. The conceptually challenging aspect is the difficulty of
explaining the correlations' origin {\em causally}, {\em
  i.e.},
according to {\em Reichenbach's principle}, stating that a
correlation between two space-time events
can stem from a {\em common cause\/} (in the common past) or a {\em direct influence\/}
from one event to the other~\cite{Reichenbach}. 
More specifically, the difficulty manifests itself when {\em
  alternatives\/}
--- hence, counterfactuals~--- are taken into account:
The argument leading up to a Bell inequality relates outcomes of
{\em alternative\/} measurements~--- only one of which can actually be realized. Does this mean that if
we drop the assumption of {\em counterfactual
  definiteness\/}~\cite{bruz}, {\em i.e.}, the requirement to
consistently 
understand counterfactual events, the paradox or strangeness disappears? 
The answer is {\em no\/}: Even 
in the ``{\em factual-only view},'' the joint properties~--- in terms of
mutual compressibility~--- of the
involved (now: fixed) pieces of information are 
such that  consequences of non-local
correlations, as understood in a common probability-calculus,
persist:
An example is the significant complexity forced upon the output given 
the input's maximal complexity plus some natural translation of
no-signaling to the  static scenario
 (see Figure~\ref{ganz}).

In the traditional, probabilistic view, a {\em Popescu-Rohrlich (PR) box\/}~\cite{pr}
gives rise to a mechanism of the following kind: Let $A$ and $B$ the 
respective input bits to the box and $X$ and $Y$ the output bits; the
(classical)
bits satisfy
\begin{align}
\label{piar}
X\oplus Y=A\cdot B\, .
\end{align}
This system is {\em no-signaling}, {\em i.e.}, the joint input-output
behavior is useless for message transmission. (Interestingly, on the
other hand, the {\em non-locality\/} of the correlation means that 
classically speaking, signaling {\em would\/} be required to {\em
  explain\/}
the behavior since shared classical information is insufficient.)
According to a result by Fine~\cite{fine}, the non-locality of the
system ({\em i.e.}, conditional distribution) $P_{XY|AB}$, which means 
that it cannot be written as a convex combination of products
$P_{X|A}\cdot P_{Y|B}$, is equivalent to the fact that there exists no
``roof distribution'' $P'_{X_0X_1Y_0Y_1}$ such that 
\[
P'_{X_iY_j} =P_{XY|A=i,B=j}
\]
for all $(i,j)\in
\{0,1\}^2$. 
In this view, non-locality means that the outputs to {\em alternative 
inputs\/} cannot consistently co\"{e}xist. The {\em counterfactual\/} nature
of this reasoning has already been pointed out by
Specker~\cite{specker}: ``In einem gewissen Sinne geh\"oren aber auch
die scholastischen Spekulationen \"uber die {\em Infuturabilien\/} hieher,
das heisst die Frage, 
ob sich die g\"ottliche Allwissenheit auch auf Ereignisse erstrecke, 
die eingetreten w\"aren, falls etwas geschehen w\"are, was nicht geschehen
ist.''~---~``In some sense, this is also related to the scholastic
speculations on the {\em infuturabili}, {\em i.e.}, the question
whether divine omniscience even extends to what would have happened
if something had happened that did not happen.'' 
Zukowski and Brukner~\cite{bruz} suggest that non-locality is to be understood in
terms of such {\em infuturabili}, called there ``counterfactual
definiteness.'' 

We intend to  challenge this view. Let us first restate
in more precise terms the counterfactual reasoning. Such reasoning is
intrinsically assuming or concluding statements of the kind that 
some piece of classical information, such as a bit $U$, {\em exists\/}
or {\em does not exist}. What does this mean? {\em Classicality\/} of information is an
idealized notion implying that it can be measured without
disturbance and that
the outcome of a measurement is always the same (which makes it clear 
this is an idealized notion requiring the classical bit to be
represented in a redundantly extended way over an {\em infinite\/} number of
degrees of freedom). It makes thus sense 
to say that a  {\em classical bit $U$ exists\/}, {\em i.e.},
has taken  a
definite value. 

In this way of speaking, Fine's theorem~\cite{fine} reads: ``The outputs
cannot {\em exist\/} before the inputs do.'' Let us make this
qualitative statement more precise. 
We assume a perfect PR box, {\em i.e.}, a system always satisfying 
$X\oplus Y=A\cdot B$. Note that this equation alone does not uniquely
determine $P_{XY|AB}$ since the marginal of~$X$, for instance, is not
determined. If, however, we additionally require {\em no-signaling}, 
then the marginals, such as $P_{X|A=0}$ or $P_{Y|B=0}$, must be perfectly 
unbiased under the assumption that all four $(X,Y)$-combinations, {\em
  i.e.},
$(0,0),(0,1),(1,0)$, and $(1,1)$, 
are possible. 
To see this, assume on the contrary that  $P_{X|A=0,B=0}(0)>1/2$. By the PR condition~\eqref{piar},
we can conclude the same for $Y$: $P_{Y|A=0,B=0}(0)>1/2$. By
no-signaling,
we also have $P_{X|A=0,B=1}(0)>1/2$. Using symmetry, and no-signaling
again, 
we obtain both $P_{X|A=1,B=1}(0)>1/2$ and $P_{Y|A=1,B=1}(0)>1/2$.
This  contradicts the PR condition~\eqref{piar} since {\em two bits which are  both
biased towards\/~$0$ cannot  differ with
certainty}. Therefore, our original assumption was wrong: The outputs 
{\em must\/} be perfectly unbiased. Altogether, this means that~$X$ as well as
$Y$ cannot exist ({\em i.e.}, take a definite value~--- actually,
there cannot 
even exist a classical value arbitrarily weakly correlated with one of
them) 
{\em before\/} for some nontrivial deterministic
function $f\/ :\/ \{0,1\}^2\rightarrow \{0,1\}$, the classical bit $f(A,B)$ exists. 
The paradoxical aspect of non-locality~--- at least if a causal
structure is  in   place~--- now consists of the fact
that {\em fresh} pieces of information {\em come to existence\/} in a
{\em spacelike-separated\/} way but that are nonetheless {\em perfectly correlated}.

\subsection{Non-Locality without Counterfactual Definiteness}

We propose an understanding of non-locality that refrains from using counterfactual
definiteness but invokes solely the data at hand, {\em i.e.},
existing in a 
single reality~\cite{pra}. 

\

\noindent
{\it Uncomputability of the outputs of a PR box.}
Let first $(a,b,x,y)$ be infinite  binary strings with
\begin{align}
\label{piri}
x_i\oplus y_i= a_i\cdot b_i\ .
\end{align}
Obviously, the intuition is that the strings stand for the inputs and
outputs of a PR box. Yet,  no dynamic meaning is attached
to the strings anymore (or to the ``box,'' for that matter) since there is
{\em no free choice of an input~--- i.e., a choice that ``could also have
  been different'' (a notion we discussed and suspect to be hard to define
  precisely in the first place)~--- and no generation of an output in function of an
  input\/}; all we have are four fixed strings satisfying the PR condition~\eqref{piri}.
However, nothing prevents us from defining this
(static) situation to be {\em no-signaling\/}: 
\begin{align}
\label{ns}
K(x\, |\, a)\approx K(x\, |\, ab)\mbox{\ \ \ and\ \  \ }K(y\, |\, b)\approx K(y\, |\, ab)\ .
\end{align}

Recall the mechanism which the maximal non-locality displayed by the
PR box enables: {\em If the inputs are not entirely fixed, then the
outputs must be completely unbiased as soon as the system is
no-signaling.}
We can now draw a similar conclusion, yet entirely within actual~---
and without having to refer  to counterfactual~--- data: 

\

\noindent
{\em If the inputs
are  incompressible and independent, and no-signaling holds,
then the outputs must be uncomputable}.

\

For a proof of this, 
let $(a,b,x,y)\in (\{0,1\}^{\bf N})^4$ with $x\oplus y=a\cdot b$
(bit-wisely), no-signaling~\eqref{ns}, and 
\[
K(a,b)\approx 2n\ ,
\]
{\em i.e.}, the ``input'' pair is  incompressible. We  conclude 
\[
K(a\cdot b\, |\, b)\approx n/2\ .
\]
Note first  that $b_i=0$ implies $a_i\cdot b_i=0$, and second that 
any further compression of $a\cdot b$, given $b$, would lead to
``structure in $(a,b)$,'' {\em i.e.},   a
possibility of describing (programming)
$a$ given $b$ in shorter than $n$ and, hence,
 $(a,b)$ in
shorter than $2n$.
Observe now
\[
K(x\, |\, b)+K(y\, |\, b)\geq K(a\cdot b\, |\, b)\ ,
\]
which implies
\begin{align}
\label{b1}
K(y\, |\, b)\geq K(a\cdot b\, |\, b)-K(x\, |\, b)\gtrsim n/2-K(x)\ .
\end{align}
On the other hand, 
\begin{align}
\label{b2}
K(y\, |\, a,b)\approx K(x\, |\, a,b)\leq K(x)\ .
\end{align}
Now, no-signaling~\eqref{ns} together with~\eqref{b1} and~\eqref{b2}
implies
\[
n/2-K(x)\lesssim K(x)\ ,
\]
and 
\[
K(x)\geq n/4 =\Theta(n)\ :
\]
The string $x$ must be uncomputable.

\

We have seen that if the pair of inputs $(a,b)$ is maximally incompressible,
then the outputs $x$ and $y$ must at least be uncomputable. This
observation raises a number of natural questions: Does a
similar 
result hold with respect to the {\em conditional\/} complexities
$K(x\, |\, a)$ and $K(y\, |\, b)$? 
With respect to {\em quantum\/} non-local 
correlations? Can we give a suitable {\em general definition\/} of
non-locality and does a similar result as the above hold with respect
to {\em any\/} non-local correlation? 
Can we strengthen and tighten our arguments to show, for instance,
that {\em uncomputable\/} inputs plus no-signaling and maximal non-locality
leads to {\em incompressibility\/} of the outputs?
What results might turn out to be 
{\em incompressibility-amplification\/}
methods. 
Let us address
these questions. 

\

\noindent
{\it Conditional uncomputability of the outputs of a PR box.}
With respect to the same assumptions as in the previous section, we
now consider the quantities $K(x\, |\, a)$ and $K(y\, |\, b)$, respectively. Note
first
\[
K(x\, |\, a)\approx 0\Leftrightarrow K(x\, |\, ab)\approx  K(y\, |\, ab)\approx 0 \Leftrightarrow K(y\, |\, b)\approx
0, 
\]
{\em i.e.}, the two expressions  vanish simultaneously. 
We show that, in fact, they  both {\em fail to be of order\/} $o(n)$.
In order to see this, assume $K(x\, |\, a)\approx 0$ and $K(y\, |\, b)\approx 0$.
Hence, there exist programs $P_n$ and $Q_n$ (both of length $o(n)$) 
for functions $f_n$ and $g_n$ with
\begin{align}
\label{fgnl}
f_n(a_n)\oplus g_n(b_n)=a_n\cdot b_n\ .
\end{align}
For fixed (families of) functions $f_n$ and $g_n$, asymptotically 
 how many $(a_n,b_n)$ can at most exist that satisfy~\eqref{fgnl}? 
The question boils down to a {\em parallel-repetition\/} analysis of
the {\em PR game\/}: A result by Raz~\cite{raz} states that when a game which cannot be won with certainty is repeated in parallel, then the success probability for all runs toghether is exponentially (in the number of repetitions) decreasing; this implies in our case that the number in question
is of order $(2-\Theta(1))^{2n}$. Therefore, the two programs $P_n$ and
$Q_n$ together with the index, of length
\[
(1-\Theta(1))2n\ ,
\] 
 of the correct pair $(a,b)$ within the list 
of length $(2-\Theta(1))^{2n}$ lead to a program, generating $(a,b)$,
that has 
length
\[
o(n)+(1-\Theta(1))2n\ ,
\] 
in contradiction to the assumption of  incompressibility of $(a,b)$.

\

\noindent
{\it Conditional uncomputability from quantum correlations.}
In the ``traditional view'' on non-locality, the PR box is an
idealization unachievable by the behavior of any quantum state. 
If it {\em did\/} exist, on the other hand, it would be a most
precious resource, {\em e.g.}, for cryptography or randomness 
amplification. The reason is that~--- as we have discussed above~--- 
under the minimal assumption that the inputs are {\em not completely 
determined}, the outputs are {\em perfectly random}, even 
given the inputs.  

Perfect PR boxes are not predicted by
quantum theory, but
sometimes, the best approximations to PR boxes that are quantum
physically achievable ($\sim 85\%$) can be used for
information-processing tasks, 
such as 
key agreement~\cite{esth}. For our application here, however, we found
this not to be the case. 
On the other hand,
 it has been shown~\cite{kent},\, \cite{colren},\, \cite{colrenamp}
that correlations which {\em are\/} achievable in the laboratory~\cite{tit}
allow for similar applications; they are based on the {\em chained
  Bell inequality\/} instead of perfect PR-type non-locality.
We show the same to hold here.

To the chained Bell inequality belongs the following idealized system: 
Let $A,B\in\{1,\ldots,m\}$ be the inputs. We assume  the
``promise''
that $B$ is congruent to $A$ or to $A+1$ modulo $m$. Given this 
promise, the outputs $X,Y\in\{0,1\}$ must satisfy
\begin{align}
\label{chain}
X\oplus Y={\chi}_{A=m,B=1}\ ,
\end{align}
where ${\chi}_{A=m,B=1}$ is the characteristic function of the event
$\{A=m,B=1\}$. 

Barrett, Hardy, and Kent~\cite{kent} showed
that if $A$ and $B$ are random, then $X$ and $Y$ must be perfectly 
unbiased if the system is no-signaling. More precisely, they were 
even able to show such a statement from the gap between the 
error probabilities of the best classical~--- $\Theta(1/m)$~---
and quantum~--- $\Theta(1/m^2)$~--- strategies for winning this
game.

In our framework, we show the following statement. 

\

\noindent
{\em Let 
$(a,b,x,y)\in (\{1,\ldots,m\}^n)^2\times (\{0,1\}^n)^2$ be such
that the promise holds, and such that
\begin{align}
\label{max}
K(a,b)\approx (\log m+1)\cdot n\ ,
\end{align}
{\em i.e.}, the string $a||b$ is maximally incompressible 
 given the promise; the system is no-signaling~\eqref{ns};
the fraction of quadruples $(a_i,b_i,x_i,y_i)$, $i=1,\ldots,n$, satisfying~\eqref{chain} 
is of order $(1-\Theta(1/m^2))n$. Then $K(x)=\Theta(n)$.}

\

Let us prove this statement. First, $K(a,b)$ being
maximal implies
\begin{align}
\label{chib}
K(\chi_{a=m,b=1}\, |\, b)\approx \frac{n}{m}\ :
\end{align}
The fractions of $1$'s in $b$ must, asymptotically,
be $1/m$ due to the string's incompressibility. If we condition on 
these positions, the string $\chi_{a=m,b=1}$ is
incompressible, since otherwise there would be the possibility of
compressing $(a,b)$.

Now, we have
\[
K(x\, |\, b)+K(y\, |\, b)+h(\Theta(1/m^2))n\gtrsim K(\chi_{a=m,b=1}\, |\, b)
\]
since one possibility for ``generating'' the string $\chi_{a=m,b=1}$, 
from position $1$ to $n$, is to generate $x_{[n]}$ and $y_{[n]}$ as well as 
the string indicating the positions where~\eqref{chain}
is 
violated, 
the complexity of the latter being at most\footnote{Here, $h$
  is the binary entropy $h(x)=-p\log p-(1-p)\log(1-p)$. Usually, $p$ is a probability, but $h$ is invoked here merely as an approximation for binomial coefficients.}\[
\log {n \choose \Theta(1/m^2)n}\approx h(\Theta(1/m^2))n\ .
\]

Let us compare this  with $1/m$: Although 
the binary entropy function has slope $\infty$ in 0, we have
\[
h(\Theta(1/m^2))<1/(3m)
\]
if $m$ is sufficiently large. To see this, observe first that the
dominant term of $h(x)$ for small $x$ is $-x\log x$, and second that
\[
c(1/m)\log(m^2/c)<1/3
\]
for $m$ sufficiently large.

Together with~\eqref{chib}, we now get
\begin{align}
\label{c1}
K(y\, |\, b)\gtrsim \frac{2n}{3m}-K(x)
\end{align}
if $m$ is chosen  sufficiently large. On the other
hand, 
\begin{align}
K(y\, |\, ab)&\lesssim K(x\, |\, ab)+h(\Theta(1/m^2))n\\
\label{c2}
&\leq K(x)+\frac{n}{3m}\ .
\end{align}

Now,~\eqref{ns}, \eqref{c1}, and~\eqref{c2} together imply 
\[
K(x)\lesssim \frac{n}{6m}=\Theta(n)\ ;
\]
in particular, $x$ must be uncomputable.

\

For any non-local behavior  characterizable by a 
condition that is always  satisfiable with entanglement, 
but not {\em without\/} this resource~| so called ``pseudo-telepathy''
games~\cite{bbt}~|, the application of Raz' {\em parallel-repetition theorem\/}
 shows 
that incompressibility of the inputs leads to 
uncomputability of at least one of the two  outputs
{\em even given the respective input}, {\em i.e.}, 
\[K(x\, |\, a)\not\approx 0\mbox{ {\em or\/} }K(y\, |\, b)\not\approx
0\ .\]
We illustrate the 
argument with the example of the {\em magic-square game\/}~\cite{ara}:
Let $(a,b,x,y)\in(\{1,2,3\}^{\bf N})^2\times (\{1,2,3,4\}^{\bf N})^2$
be the quadruple of the inputs and outputs,
respectively, and assume that the pair~$(a,b)$ is incompressible
as well as $K(x\, |\, a)\approx 0\approx K(y\, |\, b)$. Then there exist
$o(n)$-length programs $P_n$, $Q_n$ such that $x_{[n]}=P_n(a_{[n]})$ and $y_{[n]}=Q_n(b_{[n]})$. 
The parallel-repetition theorem~\cite{raz} implies that the
length of a program generating $(a_{[n]},b_{[n]})$ is, including the employed sub-routines $P_n$ and $Q_n$,
of order 
$(1-\Theta(1))\mbox{len}(a_{[n]},b_{[n]})$~--- in contradiction to the incompressibility of~$(a,b)$. 

\

\noindent
{\it An all-or-nothing flavor to the Church-Turing hypothesis.}
Our lower bound on $K(x\, |\, a)$ or on $K(y\, |\, b)$
means that if the experimenters are given access
to an incompressible number (such as $\Omega$) for choosing 
their measurement bases, then the measured photon (in a least one of
the two labs) is forced
to generate an uncomputable number as well, even given the 
string determining its basis choices. Roughly speaking, there 
is either no incompressibility at all in the world, or it is full of
it.
We can interpret that as an  all-or-nothing flavor attached
to the Church-Turing hypothesis: Either {\em no\/} physical system at all
can carry out ``beyond-Turing''  computations, or {\em even a single photon
can}.

\

\noindent
{\it General definition of (non-)locality without
  counterfactuality.}
We propose the following definition of when a no-signaling quadruple 
$(a,b,x,y)\in (\{0,1\}^{\bf N})^4$ (where $a,b$ are the ``inputs''
and $x,y$ the outputs) is {\em local}: There must exist
$\lambda\in (\{0,1\}^{\bf N})^{\bf N}$ such that 
\begin{align}
K(a,b,\lambda)& \approx K(a,b)+K(\lambda)\ ,\label{lambdainsicht}\\
K(x\, |\, a\lambda) & \approx 0\ , \mbox{\ and}\\
K(y\, |\, b\lambda) & \approx 0\ .
\end{align}

Sufficient conditions for locality are then
\[
K(a,b)\approx 0\mbox{\ \ \ or\ \ \ }K(x,y)\approx 0\ ,
\] 
since we can set $\lambda:=(x,y)$. At the other end of the scale, 
we expect that for any non-local ``system,'' the fact that $K(a,b)$ is
maximal  implies that $x$ or $y$ is conditionally  uncomputable,
given $a$ and $b$, respectively.

It is a natural question whether the given definition harmonizes with
the 
probabilistic understanding. Indeed, the latter can be seen as a special 
case of the former: If the (fixed) strings are {\em
  typical sequences\/} of a stochastic process, our non-locality
definition
implies non-locality of the corresponding  conditional  distribution. 
The reason is that a hidden variable of the distribution 
immediately
gives rise, through sampling, to
a $\lambda$ in the sense of~\eqref{lambdainsicht}.
Note, however, 
that our formalism is {\em strictly more general\/} since asymptotically, almost all strings fail to be  typical sequences of such a  process.

\section{Dropping Causality 1: Objective Thermodynamics and the Second
  Law}\label{td}

It has already been observed that the notion of 
Kolmogorov complexity can allow, in principle, for  {\em 
thermodynamics independent of probabilities or ensembles\/}:
Zurek~\cite{zurek} defines physical entropy $H_p$ to be 
\[
H_p(S):=K(M)+H(S\, |\, M)\ ,
\]
where $M$ stands for the collected data at hand while $H(S\, |\, M)$ is  the
remaining conditional
Shannon entropy of the microstate~$S$ given~$M$. 
That definition of a macrostate~($M$) is {\em subjective\/} since it 
depends on
the
{\em  available\/} data. 
How instead can  the macrostate~| and  {\em entropy}, for that matter~| 
 be defined {\em objectively\/}?
We propose 
to use
the {\em Kolmogorov sufficient statistics}~\cite{vitstat} of the
microstate: 
For any $k\in{\bf N}$, let $M_k$ be the smallest set
such that $S\in M_k$ and $K(M_k)\leq k$ hold. Let further $k_0$ be the 
value of~$k$ at which the function log-size of the set,~$\log|M_k|$, becomes linear with 
slope~$-1$. Intuitively speaking, $k_0$ is the point beyond which there is no
more ``structure'' to  exploit for describing $S$ within~$M_{k_0}$:
$S$ is a ``typical element'' of the set~$M_{k_0}$. 
We define $M(S):=M_{k_0}$ to be 
{\em $S$'s macrostate}. It~yields a program
generating $S$ of minimal length 
\[
K(S)=k_0+\log|M_{k_0}|=K(M(S))+\log|M(S)|\, .
\]
\begin{figure}[h]
	\centering
	\includegraphics[scale=1.2]{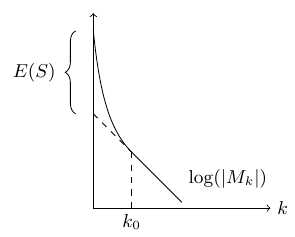}
	\caption{Kolmogorov sufficient statistics, macrostate, and
          fuel value.}
	\label{fig:6}
\end{figure}
The fuel value (as discussed in Section~\ref{fuel}) of a string~$S\in\{0,1\}^N$ 
is now related to the  macrostate $M(S)\ni S$ by
\[
E(S)\leq N-K(M(S))-\log|M(S)|
\]
(see Figure~\ref{fig:6}): Decisive is neither the  complexity
of the macrostate nor its  log-size {\em alone}, but their {\em sum}.

\begin{figure}[h]
	\centering
	\includegraphics[scale=1]{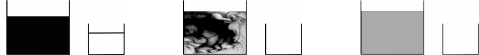}
	\caption{Coffee and milk.}
	\label{fig:4}
\end{figure}

A notion
defined in a related way is the {\em
  sophistication\/} 
or {\em interestingness} as discussed by Aaronson~\cite{aare} 
investigating 
the process where milk is poured
into coffee (see Figure~\ref{fig:4}). 
Whereas the initial and final
states are ``simple'' and ``uninteresting,'' the intermediate
(non-equilibrium)
states display a rich structure; here, the
sophistication~| and also $K(M)$ for our macrostate $M$~| 
becomes maximal. 

\begin{figure}[h]
	\centering
	\includegraphics[scale=1]{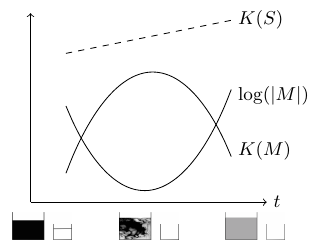}
	\caption{The complexity and the size of the macrostate.}
	\label{fig:5}
\end{figure}

During the process under consideration, neither 
the macrostate's complexity nor its size is
monotonic in time: Whereas $K(M)$ has a {\em maximum\/} in the non-equilibrium phase of 
the process, $\log|M|$ has a {\em minimum\/} there (see
Figure~\ref{fig:5}). 

On the other hand, the complexity of the {\em microstate},
\[
K(S)=K(M)+\log|M|\ ,
\]
is  a candidate for a (essentially) monotonically nondecreasing quantity: Is this
the {\em second law of thermodynamics\/} in that
view? This law, which claims a certain quantity to be (essentially)
monotonic in time,  is by many believed to be the origin of our 
ability to distinguish the future from the past. 

\ 

\noindent
{\it The second law, traditional view.}
Let a  closed system be in a thermodynamical equilibrium state of entropy
$S_1$ at time~$t_1$. Assume that the system evolves to another equilibrium
state, of entropy $S_2$, at some fixed later time $t_2>t_1$. Then, for
$s>0$,
\[
{\rm Prob}[S_1-S_2\geq sk\ln 2]=2^{-s}\ .
\]

\

It is a rare example~| outside quantum theory
|  of a physical ``law'' holding  only with some probability.

\

\noindent
{\em Is there an underlying 
fact in the form of a property of an evolution holding with
certainty and also for all intermediate states?}

\

 Clearly, that fact
would not talk about the {\em coarse-grained\/} behavior of the 
system, which we have seen in the discussed example to be 
{\em non\/}-monotonic in time. If, however, we consider the
{\em micro\/}state, then {\em logical reversibility\/}~| meaning that the
past 
can be computed step by step from the future (not necessarily {\em
  vice versa\/})~| is a good candidate:
Indeed, also Landauer's principle links the second law to logical
irreversibility. A logically reversible evolution is 
potentially asymmetric in time if the backward direction is 
{\em not\/} logically reversible.

In the spirit of the {\em Church-Turing hypothesis}, we see the state
of a closed system in question as a finite binary string and its
evolution (through discretized time) as being computed by a universal Turing machine.

\ 

\noindent
{\it The second law, revisited.}
The evolution of a closed system 
is {\em logically reversible\/} and 
the past at time $t_1$ can be computed from the future at time $t_2\, (>t_1)$
by a constant-length 
program on a Turing machine. 

\

It is somewhat ironic that this view of the 
second law puts forward the {\em reversibility\/} of the computation,
whereas the law
is usually linked to the  opposite:
{\em irreversibility}.
A consequence of the law is that the decrease of the  Kolmogorov complexity of the
string
encoding the system's state is limited.

\ 

\noindent
{\it Consequence  of the second law, revisited.}
Let $x_1$ and $x_2$ be the 
contents of a reversible Turing machine's tape at times 
$t_1<t_2$.
Then
\[
K(x_1)\leq K(x_2)+\Theta(\log(t_2-t_1))\ .
\]

\begin{figure}[h]
	\centering
	\includegraphics[scale=1]{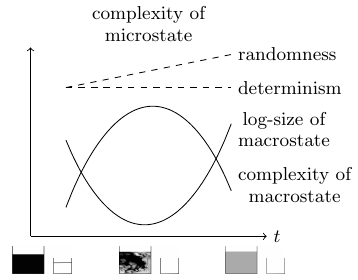}
	\caption{Randomness vs.\ determinism.}
	\label{figueres}
\end{figure}

If the Turing machine is {\em deterministic}, the complexity 
increases at most logarithmically in time. On the other hand,
this growth can of course be arbitrarily faster for {\em
  probabilistic\/}
machines.
Turned around, Kolmogorov complexity can yield an 
{\em intrinsic\/} criterion 
for the distinction between determinism and indeterminism (see
Figure~\ref{figueres}). In the case of randomness, a strong 
asymmetry and an objective arrow of time can arise. 
A context-free definition of randomness (or free will for that matter)
has the advantage not to depend on the ``possibility that something could have
been different from how it was,'' a metaphysical condition we
came to prefer to avoid.

\

\noindent
{\it The traditional second law from complexity increase.}
It is natural to ask what the connection between logical 
reversibility and complexity on one side and the traditional 
second law on the other is. We show that the latter emerges from increasing 
complexity | including the exponential error probabilities.

Let $x_1$ and $x_2$ be the microstates of a closed system at times
$t_1<t_2$ with $K(x_2)\geq K(x_1)$.
If the macrostates $M_1$ and $M_2$ of $x_1$ and $x_2$, respectively, 
have {\em small Kolmogorov complexity\/} (such as traditional thermodynamical
equilibrium states characterized by global parameters like volume, 
temperature, pressure, etc.), then
\[
|M_1|\lesssim |M_2|\ :
\]
If the macrostates are simple, then their size is non-decreasing. 
Note that this law is still compatible with 
the exponentially small
error probability ($2^{-N}$) in the traditional view of the
second law for a spontaneous immediate drop of entropy by $\Theta(n)$: 
The gap opens
when the simple thermodynamical equilibrium macrostate of a given
microstate differs from our macrostate defined through the
Kolmogorov statistics.  This can occur if, say, the positions and momenta
of the molecules of 
some (innocent-, {\em i.e.}, general-looking) gas 
encode, {\em e.g.},~$\pi$ and have essentially zero complexity.

\

We can now finish up  by closing a logical circle. 
We have started from the converse of
Landauer's 
principle, went through work extraction and ended up with a 
complexity-theoretic view of the second law: We have returned back to our starting point.

\

\noindent
{\it Landauer's principle, revisited.}
The (immediate) transformation of a string $S$ to the $0$-string of the same
length
requires free energy at least \[K(S)kT\ln2\ ,\] which is then dissipated
as heat to the environment. For every concrete lossless compression 
algorithm $C$, 
\[
{\rm len}(C(S)) kT\ln2+\Theta(1)\ ,
\]
is, on the other hand, an upper bound on the required free energy.

\

Finally,
Landauer's principle
can be combined with its converse and generalized 
as follows.

\

 \noindent
{\it Generalized Landauer's principle.}
Let $A$ and $B$ two bit strings of the same length. The (immediate) transformation 
from $A$ to $B$ costs at least
\begin{align}
(K(A)-K(B))kT\ln 2 \label{thermo}
\end{align}
free energy, or it releases at most the absolute
value of~\eqref{thermo} if this  is negative. 

\

\begin{figure}[h]
	\centering
	\includegraphics[scale=1]{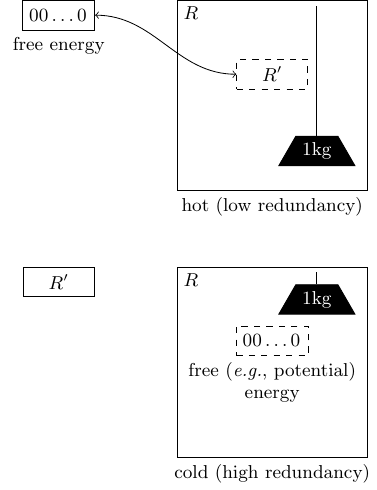}
	\caption{Work extraction and Landauer's principle in the view
          of ``the church of the larger bit string.''}
	\label{wheel}
\end{figure}

If the Turing machine is a closed physical system, then this principle 
reduces to the complexity-non-decrease stated above. This suggests that the physical system possibly {\em simulated\/} by the
machine~| 
in the spirit of the Church-Turing hypothesis~| 
also follows the second law ({\em e.g.}, since it is a closed system
as well). 
The fading boundaries between what the machine {\em is\/} and what 
is {\em simulated\/} by it are in accordance with Wheeler's~\cite{wheeler} ``it from
bit:'' {\em Every ``it''~| every particle, every field of force, even
  the spacetime continuum itself~| 
derives its function, its meaning, its very existence entirely [\ldots] from the apparatus-elicited 
answers to yes or no questions, binary choices, ``bits.''}
If we try to follow the lines of such a view further, we may model the
environment as a binary string $R$ as well.
The goal is a unified discourse avoiding to speak 
about complexity with respect to one system and about free energy, heat,
and temperature to the other.
 The transformation
addressed by 
Landauer's principle and its converse then looks as in
Figure~\ref{wheel}: The low-complexity zero-string can be swapped with 
``complexity'' in the environment which in consequence
becomes more redundant, {\em i.e.}, 
 cools down
but receives
free energy, for instance in the form of a
weight having been lifted.

\section{Dropping Causality 2: Space-Time from Complexity}\label{spacom}

\subsection{Information and Space-Time}

If, motivated by the above, we choose to regard {\em information\/} as being
more fundamental than space and time, how can the latter be imagined
to emerge from the former? Can such a causal structure be understood
to be of  {\em logical\/} 
rather than {\em physical\/} nature? In other words, is it more
accurate to imagine causal relations to be a property of logical
rather than physical spaces~\cite{witt}?
We address these questions here, continuing to avoid speaking about ``what could have 
been different,'' {\it i.e.}, the counterfactual viewpoint.

In Section~\ref{td}, an arrow of time has emerged under the assumption
of (uni-directional) logical reversibility. Here, we refine the same idea in an attempt
to derive a causal structure based on the  principle that any point carries
complete information about its space-time past.

\subsection{Causal Structures}
\label{sec:causalstructures}
Let us start  with a finite set ${\cal C}$ of  strings on which we 
would like to find a causal structure arising {\em from inside}, {\em
  i.e.}, from the
properties of, and relations between,  these strings. 
The intuition is that an $x\in{\cal C}$
encodes the totality of momentary local physical reality in a ``point,'' {\em i.e.}, parameters such as mass, charge, electric
and magnetic field density.

Let ${\cal C}\subset\{0,1\}^{\bf N}$ be  finite.
We define the following  order relation on ${\cal C}$:\footnote{In 
this section, conditional complexities are understood as follows:
In $K(x\, |\, y)$, 
for instance, the condition $y$ is assumed to be the full (infinite)
string, 
whereas the asymptotic process runs over
  $x_{[n]}$. The reason is that very insignificant bits of $y$
  (intuitively: the present) can be
  in relation to bits of $x$ (the past) of much higher significance.
The past does not disappear, but it fades.}\[
x\preceq y :\Longleftrightarrow
K(x\, |\, y)\approx 0\ .
\]
We say that $x$ is a {\em cause\/} of $y$, and that $y$ is an {\em
  effect\/} of~$x$.
So, $y$ is in $x$'s future
exactly if $y$ contains
the entire information about $x$; no information is ever lost. 
The intuition is that {\em any\/} ``change'' in the cause affects {\em each one\/} of its
effects~--- if sufficient precision is taken into account.
We write $x\doteq y$ if $x\preceq y$ as well as $x\succeq y$ hold. 
If $x\not\preceq y$ and $x\not\succeq y$, we write 
$x\not\preceq\not\succeq y$ and call $x$ and~$y$ {\em spacelike
separated}. 
We call the pair $({\cal C},\preceq)$ a 
{\em  causal structure}. 

For a set $\{x_i\}\subseteq {\cal C}$ and $y\in {\cal C}$, we say that
$y$ is the {\em first common
effect\/} of the $x_i$ if it is the least upper bound: $x_i\preceq y$
holds for all $x_i$, and  for any $z$ 
with $x_i\preceq z$ for all $x_i$, also $y\preceq z$ holds. The notion
of {\em last common cause\/} is defined analogously.
A~minimum (maximum) of $({\cal C},\preceq)$ is called {\em without
  cause\/}
({\em without effect\/}). 
 If ${\cal C}$ has
a smallest (greatest) element, this is called {\em big bang\/} ({\em big
  crunch\/}). 

We call a causal structure {\em deterministic\/} if, intuitively,
every $y$ which is not without cause is completely
determined by all its causes. Formally, for some $y\in {\cal C}$, let $\{x_i\}$ be
the set of all $x_i\in {\cal C}$ such that $x_i\preceq y$ holds. Then
we must have
\[
K(y\, |\, x_1,x_2,\ldots)\approx 0\ .
\]
Otherwise, ${\cal C}$ is called {\em probabilistic}.

\
\\ 
\

\subsection{The Emergence of Space-Time}

Observe first that {\em every deterministic causal structure which has a
big bang is trivial\/}: We have
\[
x\doteq y\mbox{\ \ for all\ \ }x,y\in{\cal C}\ .
\]
This can be seen as follows. Let $b$ be the big bang, {\em i.e.}, 
$b\preceq x$ for all $x\in{\cal C}$. On the other hand,
$K(x\, |\, z_i)\approx 0$
if $\{z_i\}$ is the set of predecessors of $x$. Since the same is true 
for each of the $z_i$, we can continue this process and, ultimately,
end up with only $b$: $K(x\, |\, b)\approx 0$, {\em i.e.}, $x\preceq b$, and
thus $b\doteq x$ for all $x\in{\cal C}$. 
In this case, we obviously cannot expect to be able to explain space-time. (Note, however, that 
there can still exist deterministic ${\cal C}$'s~--- without big bang~--- with non-trivial 
structure.)
However, the world as it presents itself to us~--- with {\em both\/} big bang {\em
  and\/} arrow of time~--- seems to
direct us
away from determinism (in support of~\cite{gisin}).

The situation is very different in {\em probabilistic\/} causal
structures: Here, the partial order relation $\preceq$ gives rise 
to a non-trivial picture of causal relations and, ideally, causal space-time 
including the arrow of time. 
Obviously, the resulting structure depends crucially on the set 
${\cal C}$. 
Challenging open problems are to understand the relationship between
sets of strings and causal structures: Can every partially ordered set 
be implemented by a suitable set of strings? 
What is the property of a set of strings that gives rise to the
``usual''
space-time of relativistic light-cones?

Is it helpful to introduce a {\em metric\/} instead of just an
order relation? As a first step, it appears natural to define 
$K(y\, |\, x)$ as the {\em distance of $x$ from the set of effects of $y$}.
In case $y$ is an effect of $x$, this quantity intuitively measures 
the {\em time\/} by which $x$ happens {\em before\/} $y$.

Generally in such a model, 
what is a ``second law,'' and under what condition does it hold?
Can it~--- and the arrow of
time~--- be compatible even with determinism (as long as there is no big bang)?

What singles out the sets
displaying quantum non-local correlations as observed in the lab? (What
is the significance of Tsirelson's bound in the picture?)

\section{Dropping Causality 3: Preserving Logical Consistency}
A recent framework for quantum~\cite{ocb} and classical~\cite{njp} correlations without causal order is based on {\em local\/} assumptions only.
These are the local validity of quantum or classical probability theory, that laboratories are closed (parties can only interact through the environment), and that the probabilities of the outcomes are linear in the choice of local operation.
The {\em global\/} assumption of a {\em fixed global causal order\/} is replaced by the assumption of {\em logical consistency\/}: All probabilities must be non-negative and sum up to~$1$.
Some correlations~--- termed {\em non-causal\/}~--- that can be
obtained in this picture cannot arise from global quantum or classical
probability theory.
Similarly to the discovery of non-local correlations that showed the existence of a
world between the local and the signaling; in a similar sense, we
discuss here  {\em a territory that lies between what is causal and what is
logically inconsistent: It is not empty}.

In the spirit of Section~\ref{nl}, where we studied the consequences of non-locality, we show that the results from non-causal correlations carry over to the picture of (conditional) compressibility of bit strings, where we do not employ  probabilities, but consider {\em actual\/} data only.
In that sense, these are the {\em non-counterfactual\/} versions of  results on non-causal correlations.

\subsection{Operational Definition of Causal Relations}
We define causal relations operationally, where we use the notion of parties.
A {\em party\/} can be thought of as a system, laboratory, or an experimenter, performing an operation.
In the traditional view, the choice of operation is represented by {\em randomness}, and thus by a probability distribution.
Here, in contrast, we refrain from this counterfactual approach (probabilities), and consider {\em actual\/}~--- as opposed to {\em potential\/}~--- choices only.
The traditional view with probabilities has a {\em dynamic\/}
character: Systems undergo (randomized) evolutions.
Like in Section~\ref{nl}, we obtain a {\em static\/} situation if  we consider actual data only.
All statements are formulated with bit strings and {\em relations\/}
between these  strings modeling 
the ``operations.''

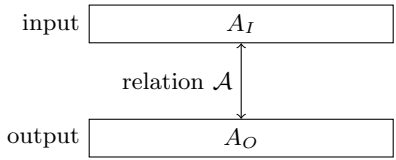
\begin{figure}[h]
	\centering
	\begin{tikzpicture}
		\node[draw,rectangle,minimum width=4cm,minimum height=0.5cm,label=left:input] (I) {$A_I$};
		\node[draw,rectangle,minimum width=4cm,minimum height=0.5cm,label=left:output,below=1cm of I] (O) {$A_O$};
		\draw[<->] (I) -- (O) node[midway,left] (l) {relation~$\mathcal{A}$};
	\end{tikzpicture}
	\caption{A party~$A$ as two bit strings~$A_I$ (input) and~$A_O$ (output) that satisfy some relation~$\mathcal{A}$.}
	\label{fig:party}
\end{figure}

A party~$A$ is modeled by two bit strings~$A_I$ and~$A_O$.
We restrict ourselves to pairs of bit strings that satisfy some relation~$\mathcal{A}$.
Within a party, we assume a fixed causal structure ($A_I$ precedes~$A_O$) (see Figure~\ref{fig:party}).
The relation~$\mathcal{A}$ is called {\em local operation of~$A$}, the
string~$A_I$ is called {\em input to~$A$\/}, and~$A_O$ is  {\em $A$'s output}.
If we have more than one party, we consider only those input and
output bit strings that satisfy some 
{\em global relation}.
These relations are, as in Section~\ref{nl}, to be understood to act
{\em locally\/}
on the involved strings: A relation involves only a finite number of
instances (bit positions), and it is repeated~$n\, (\rightarrow\infty)$
times for obtaining the global relation.

For two parties~$A$ and~$B$, we say that~$A$ is in the {\em causal
  past\/} of~$B$, $A\preceq B$, if and only if
\begin{align}\label{causalcondi}
	K(B_I\, |\, A_O)\not \approx K(B_I)\not\approx 0
	\,.
\end{align}
Intuitively,~$A$ is in the causal past of~$B$ if and only if~$B$'s
input is {\em uncomputable\/}~--- otherwise~$B$ could simply obtain it
herself~--- and better {\em compressible\/} with~$A$'s output than
without, {\em i.e.}, the two strings depend on each other.
Expressed according to intuitive dynamic thinking, the definition
means that~$A$ 
is in the causal past of~$B$ if and only if~$B$ {\em learns\/} parts of an incompressible string from~$A$.
The causal relation among parties defined here is extended
straight-forwardly 
to the scenario where {\em one or more\/} parties are in the causal past of {\em one or more\/} parties.

This definition is different from the one proposed in Section~\ref{spacom}.
There, a  string~$x$ is said to be the {\em cause\/} of another  string~$y$ (the {\em effect}) if and only if~$K(x\,|\,y)\approx 0$.
The intuition there is {\em logical reversibility\/}: Future events
contain all information about past events, no information is ever
lost, and~$x$ and~$y$ are understood to encode complete physical reality in some space-time point.  
In contrast, the definition here only relates pieces of information
chosen and processed by the parties: If one party's input depends on
another's output, then she is in the causal future of the latter. 
(Reversibility, the central notion in
Section~\ref{spacom},  
does not play a role here.)
Since the strings now just correspond to the pieces of information manipulated by the parties, we cannot simply  define freeness  as an attribute of complexity. Instead, we {\em postulate\/} the output  strings to be free. 

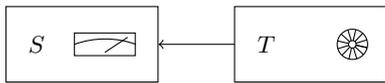
\begin{figure}[h]
	\centering
	\begin{tikzpicture}
		\node[draw,shape=rectangle,minimum width=2cm,minimum height=1cm] (S) {};
		\node[right=0.3cm of S.west,inner sep=0pt,outer sep=0pt] (St) {$S$};
		\node[left=0.3cm of S.east,draw,shape=rectangle,minimum width=0.8cm,minimum height=0.3cm] (M) {};
		\draw (M.east) arc (70:110:1.18);
		\draw[-] (M.south)++(0cm,0.05cm) -- ++(0.3cm,0.2cm);
		\node[draw,shape=rectangle,minimum width=2cm,minimum height=1cm,right=of S,inner sep=0pt,outer sep=0pt] (T) {};
		\node[right=0.3cm of T.west,inner sep=0pt,outer sep=0pt] (Tt) {$T$};
		\node[left=0.4cm of T.east,draw,shape=circle,inner sep=0pt,outer sep=0pt,minimum size=0.1cm] (K) {};
		\node[draw,shape=circle,inner sep=0pt,outer sep=0pt,minimum size=0.4cm] (K2) at (K) {};
		\foreach \angle in {
			-13+0*30,
			-13+1*30,
			-13+2*30,
			-13+3*30,
			-13+4*30,
			-13+5*30,
			-13+6*30,
			-13+7*30,
			-13+8*30,
			-13+9*30,
			-13+10*30,
			-13+11*30,
			-13+12*30
		}
		{
			\draw (K.center)++(\angle:0.05cm) -- +(\angle:0.15cm);
		}
		\draw[->] (T.west) -- (S.east);
	\end{tikzpicture}
	\caption{If a variable~$T$ is correlated to another variable~$S$, and~$T$ is {\em free\/} but~$S$ is {\em not}, then~$T$ is in the causal past of~$S$.}
	\label{fig:free}
\end{figure}

The rationale of Definition~\eqref{causalcondi}
above is similar to the one we propose in~\cite{aemin} for the probability picture.
There,~$A\preceq B$ holds if and only if both random variables~$A$ and~$B$ are {\em correlated\/} and~$A$ is postulated {\em free}.
The motivation is to define causal relations {\em based\/} on freeness, and not the other way around (see Figure~\ref{fig:free}).
Intuitively, if you flip a switch that is correlated to a light bulb, then flipping the switch is in the causal past of the light turning on or off~--- the definition of a causal relation relies on what we call {\em free\/} (the switch in this case).
Such a definition based on postulated freeness is similar to the interventionist's approach to causality, see {\it e.g.},~\cite{Woodward}.
In the approach studied here, the analog to {\em correlation\/} is {\em dependence}.
The distinction between {\em free\/} and {\em not free\/} variables is done in the same way by distinguishing between {\em input\/} and {\em output\/} bit strings.

\subsection{Causal Scenario}
\begin{figure}[h]
	\centering
	\subfloat[\label{fig:causalscenario}]{
		\includegraphics{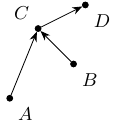}
	}
	\qquad
	\subfloat[\label{fig:noncausalscenario}]{
		\includegraphics{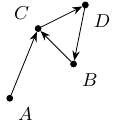}
	}
	\caption{(a) Example of a causal scenario among four parties
          with~$(A,B)\preceq C$ and~$C\preceq D$. (b) Example of a
          non-causal scenario with~$(A,B) \preceq C$,~$C\preceq D$,
          and~$D\preceq B$. Arrows point into the direction of the causal future.}
	\label{fig:scenarios}
\end{figure}

{\em Causal scenarios\/} describe input and output  strings of the
parties where the resulting causal relations reflect a {\em partial ordering\/} of the parties (see Figure~\ref{fig:causalscenario}).\footnote{Transitivity arises from the assumption of a fixed causal structure within a party, where the input is causally prior to the output.}
In the most general case, the partial ordering among the parties of a
set~$S$, who are all in the causal future of some other
party~$A\not\in S$, {\it i.e.},~for all~$B\in S:A\preceq B$, can
depend on
({\em i.e.}, satisfy some relation with)
the bit strings of~$A$~\cite{isit},~\cite{christina}.
A causal scenario, in particular, implies that {\em at least one party is not in the causal future of some other parties}.
If no partial ordering of the parties arises, then the scenario is called {\em causal\/} (see Figure~\ref{fig:noncausalscenario}).

A trivial example of a causal scenario is a communication channel over which a bit is perfectly transmitted from a party to another.
This channel, formulated as a global relation, is~$f(x,y)=(0,x)$, with~$x,y\in\{0,1\}$, and where the first bit belongs to~$A$ (sender) and the second to~$B$ (receiver) (see Figure~\ref{fig:channel}).
Consider the~$n\, (\rightarrow\infty)$-fold sequential repetition of this global relation, and assume that both output bit strings are incompressible and independent:~$K(A_O,B_O)\approx 2n$.
The bit string~$A_I$ is~$(0,0,0,\dots)$ according to the global relation.
In contrast,~$B_I$ is equal to~$A_O$.
Since~$K(B_I)\approx n$ and~$K(B_I\, |\, A_O)\approx 0$, the causal
relation~$A\preceq B$ holds, restating that~$A$ is in the causal past of~$B$.
Conversely,~$K(A_I)\approx 0$ and, therefore,~$A\not\succeq B$: The receiver is {\em not\/} in the causal future of the sender.

\subsection{Non-Causal Scenario}
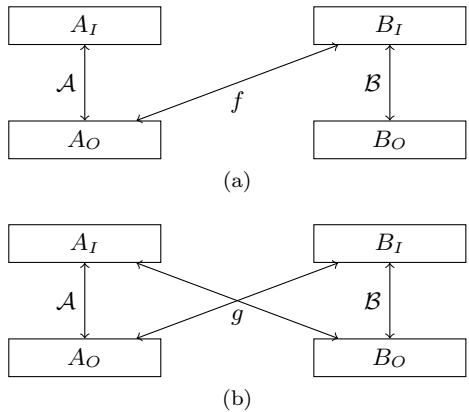
\begin{figure}[h]
	\centering
	\subfloat[\label{fig:channel}]{
		\begin{tikzpicture}
			\node[draw,rectangle,minimum width=2cm,minimum height=0.5cm] (AI) {$A_I$};
			\node[draw,rectangle,minimum width=2cm,minimum height=0.5cm,below=1cm of AI] (AO) {$A_O$};
			\node[draw,rectangle,minimum width=2cm,minimum height=0.5cm,right=2cm of AI] (BI) {$B_I$};
			\node[draw,rectangle,minimum width=2cm,minimum height=0.5cm,right=2cm of AO] (BO) {$B_O$};
			\draw[<->] (AI) -- (AO) node[midway,left] (A) {$\mathcal{A}$};
			\draw[<->] (BI) -- (BO) node[midway,left] (B) {$\mathcal{B}$};
			\draw[<->] (AO) -- (BI) node[midway,below] (R) {$f$};
		\end{tikzpicture}
	}

	\subfloat[\label{fig:2wayprocess}]{
		\begin{tikzpicture}
			\node[draw,rectangle,minimum width=2cm,minimum height=0.5cm] (AI) {$A_I$};
			\node[draw,rectangle,minimum width=2cm,minimum height=0.5cm,below=1cm of AI] (AO) {$A_O$};
			\node[draw,rectangle,minimum width=2cm,minimum height=0.5cm,right=2cm of AI] (BI) {$B_I$};
			\node[draw,rectangle,minimum width=2cm,minimum height=0.5cm,right=2cm of AO] (BO) {$B_O$};
			\draw[<->] (AI) -- (AO) node[midway,left] (A) {$\mathcal{A}$};
			\draw[<->] (BI) -- (BO) node[midway,left] (B) {$\mathcal{B}$};
			\draw[<->] (AI) -- (BO);
			\draw[<->] (AO) -- (BI) node[midway,below] (R) {$g$};
		\end{tikzpicture}
	}
	\caption{(a) The global relation~$f$ describes a channel from~$A$ to~$B$. (b) The input to party~$A$ is, as defined by the global relation~$g$, identical to the output from party~$B$, and the input to party~$B$ is identical to the output from party~$A$.}
	\label{fig:globalrelations}
\end{figure}

Consider the global relation
\begin{align}
	g(x,y)=(y,x)
	\label{eq:2wayprocess}
	\,,
\end{align}
which describes a {\em two-way channel\/}:~$A$'s output is equal to~$B$'s input and~$B$'s output is equal to~$A$'s input (see Figure~\ref{fig:2wayprocess}).
This global relation can describe a non-causal scenario.
If~$K(A_O,B_O)\approx 2n$, then indeed, the causal relations that we obtain are~$A\preceq B$ and~$B\preceq A$.
What we want to underline here is that for {\em this particular choice\/} of local operations of the parties, input bit strings that are consistent with the relation~\eqref{eq:2wayprocess} exist.
In stark contrast, if we fix the local operations of the parties to be~$A_O=\overline{A_I}$ (the output equals the bit-wise flipped input) for party~$A$ and~$B_O =B_I$ for party~$B$,
then {\em no choice of inputs\/}~$A_I$ and~$B_I$ satisfies the desired global relation~\eqref{eq:2wayprocess}.
This inconsistency is also known as the {\em grandfather antinomy}.
If no satisfying input and output  strings exist, then we say that the global relation is {\em inconsistent with respect to the local operations}.
Otherwise, the global relation is {\em consistent with respect to the local operations}.

\begin{figure}[h]
	\centering
	\includegraphics{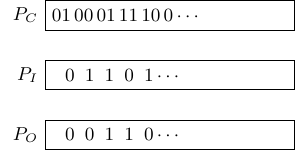}
	\caption{Example of input, output, and~$P_C$ of a party~$P$ that satisfy the relation defined by Eq.~(15).}
	\label{fig:C}
\end{figure}

For studying bit-wise global relations, {\it i.e.}, global relations
that relate single output bits with single input bits, that are consistent {\em regardless\/} the local operations, we set the local operation to incorporate all possible operations on bits.
These are the constants~$0$ and~$1$ as well as the identity and bit-flip operations.
The parties additionally hold  incompressible and independent  strings
that define which of these four relations is in place at a given bit position.
For party~$P$, let this additional bit string be~$P_C$.
Formally, if we have~$k$ parties~$A,B,C,\dots$, then
\begin{align}
	K(A_C,B_C,C_C,\dots)&\approx kn
	\,.
\end{align}
The local operation of a party~$P$ is
\begin{align}
	P_O^{(i)}=
	\begin{cases}
		0&\text{if }(P_C^{(2i)},P_C^{(2i+1)})=(0,0)\,,\\
		1&\text{if }(P_C^{(2i)},P_C^{(2i+1)})=(1,1)\,,\\
		P_I^{(i)}&\text{if }(P_C^{(2i)},P_C^{(2i+1)})=(0,1)\,,\\
		P_I^{(i)}\oplus 1&\text{if }(P_C^{(2i)},P_C^{(2i+1)})=(1,0)\,,
	\end{cases}
	\label{eq:rule}
\end{align}
where a superscript~$(i)$ selects the~$i$-th bit of a  string.
Depending on pairs of bits on~$P_C$, the relation~\eqref{eq:rule}
states that a given output bit is either equal to~$0$ or~$1$ 
or equal to or different from the corresponding input bit.
An example is presented in Figure~\ref{fig:C}.
Since all pairs of bits appear equally often in~$P_C$ (asymptotically
speaking), in half  the cases bits of the 
output  string are identical to bits of the (incompressible)  string~$P_C$ in the respective positions.
Thus, the output satisfies~$K(P_O)=\Theta(n)$.
We call the local operation of Eq.~\eqref{eq:rule} of a party {\em universal local operation}.
If a global relation is consistent with respect to universal local operations, then we call it {\em logically consistent}.
If we consider {\em all\/} (bit-wise) operations, the global relation~\eqref{eq:2wayprocess} becomes {\em inconsistent\/}:
No input and output  strings exist that satisfy the desired global relation~\eqref{eq:2wayprocess}.
To see this, note that since we are in the asymptotic case, there exist  positions~$i$ where the relation of~$A$ states that the~$i$th output bit is equal to the~$i$th input bit, and the relation of~$B$ states that the~$i$th output bit is equal to the negated~$i$th input bit, which results in a contradiction~--- the global relation {\em cannot\/} be satisfied.
In more detail, there exists an~$i$ such that the bit string~$A_C$ contains the pair~$(0,1)$ at position~$2i$, and such that the bit string~$B_C$ contains the pair~$(1,0)$ at the same position:
\begin{align}
	A_C&=(\dots,0,1,\dots)\,,\\
	B_C&=(\dots,1,0,\dots)
	\,.
\end{align}
On the one hand, the input to~$A$ has a value~$a$ on the~$i$th position, and, because of~$A_C$, the same value is on the~$i$th position of~$A_O$:
\begin{align}
	A_I&=(\dots,a,\dots)\,,\\
	A_O&=(\dots,a,\dots)
	\,.
\end{align}
The input and output bit strings of~$B$, on the other hand, must, due to~$B_C$, have opposite bits on the~$i$th position:
\begin{align}
	B_I&=(\dots,b,\dots)\,,\\
	B_O&=(\dots,b\oplus 1,\dots)
	\,.
\end{align}
A contradiction arises: No choice of bits~$a$ and~$b$ exist that satisfy the global relation~\eqref{eq:2wayprocess}:
The global relation~\eqref{eq:2wayprocess}, which is depicted in Figure~\ref{fig:2wayprocess}, is {\em logically inconsistent}.

\begin{figure}[h]
        \centering
        \begin{tikzpicture}
                \def\r{1}
                \def\d{20}
                \def\dx{2.1}
                \def\hd{.5}

                \draw[->] (-\dx,0)++(90-\d:\r) arc (90-\d:-30+\d:\r);
                \draw[->] (-\dx,0)++(-30-\d:\r) arc (-30-\d:-150+\d:\r);
                \draw[->] (-\dx,0)++(-150-\d:\r) arc (-150-\d:-270+\d:\r);
                \draw (-\dx,0)++(90:\r) node (A) {$A$};
                \draw (-\dx,0)++(-30:\r) node (B) {$B$};
                \draw (-\dx,0)++(-150:\r) node (C) {$C$};

                \draw[<-] (\dx,0)++(90-\d:\r) arc (90-\d:-30+\d:\r);
                \draw[<-] (\dx,0)++(-30-\d:\r) arc (-30-\d:-150+\d:\r);
                \draw[<-] (\dx,0)++(-150-\d:\r) arc (-150-\d:-270+\d:\r);
                \draw (\dx,0)++(90:\r) node (A2) {$A$};
                \draw (\dx,0)++(-30:\r) node (B2) {$B$};
                \draw (\dx,0)++(-150:\r) node (C2) {$C$};
                \draw (\dx,0)++(90-60:\r+.3) node (n1) {$\oplus 1$};
                \draw (\dx,0)++(-30-60:\r+.3) node (n2) {$\oplus 1$};
                \draw (\dx,0)++(-150-60:\r+.3) node (n3) {$\oplus 1$};

                \draw (-\dx-\r-\hd,\r+0.3) node (half1) {$\mathrm{maj}(a,b,c)=0$};
                \draw (\dx-\r-\hd,\r+0.3) node (half2) {$\mathrm{maj}(a,b,c)=1$};
        \end{tikzpicture}
        \caption{The left channel is chosen if the majority of the output bits is~$0$; otherwise, the right channel is chosen.}
        \label{fig:maj}
\end{figure}
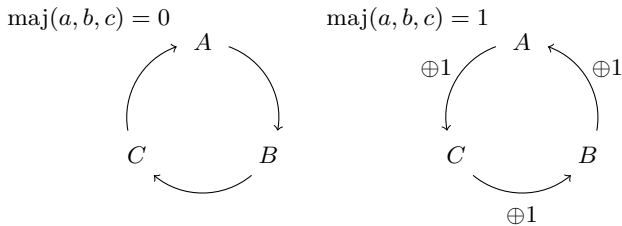

We show that there exist {\em logically consistent global relations that are non-causal\/}~\cite{aemin}.
Suppose we are given three parties~$A$,~$B$, and~$C$ with universal local operations.
There exist global relations where the input to any party is a function of the outputs from the remaining two parties.
An example~\cite{njp} of such a global relation is
\begin{align}
	x=\neg b \wedge c\,,\quad
	y=a\wedge \neg c\,,\quad
	z=\neg a \wedge b
	\,,
	\label{eq:noncausal}
\end{align}
where all variables represent bits, and where~$x,y,z$ is the input to~$A,B,C$ and~$a,b,c$ is the output from~$A,B,C$, respectively.
This global relation can be understood as follows: Depending on the {\em majority\/} of the output bits, the relation {\em either\/} describes the identity channel from~$A$ to~$B$ to~$C$, and back to~$A$, {\em or\/} it describes the bit-flip channel from~$A$ to~$C$ to~$B$, and back to~$A$ (see Figure~\ref{fig:maj}).
We study the causal relations that emerge from~$n\, (\rightarrow\infty)$ sequential repetitions of this global relation, {\it i.e.}, infinite  strings that satisfy the global relation~\eqref{eq:noncausal}.
The input to party~$A$ is uncomputable, even~$K(A_I)\not\approx 0$,
because some bit positions of the outputs from~$B$ and~$C$ are uncomputable.
Yet, the outputs from~$B$ and~$C$ {\em completely\/} determine the input to~$A$, {\it i.e.},~$K(A_I\, |\, B_O,C_O)\approx 0$.
Therefore, the causal relation~$(B,C)\preceq A$ holds.
Due to symmetry, the causal relations~$(A,C)\preceq B$ and~$(A,B)\preceq C$ hold as well.
All together imply that {\em every\/} party is in the causal future of some other parties~--- the scenario is {\em non-causal}.
On the other hand, it is {\em logically consistent\/}: 
There exist input and output bit strings that satisfy the global relation~\eqref{eq:noncausal} at every bit-position.

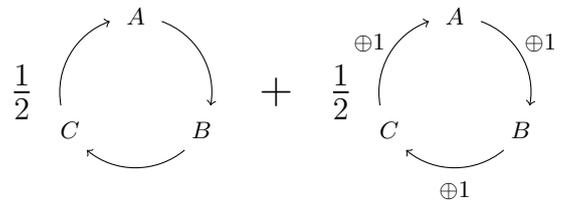
\begin{figure}[h]
        \centering
        \begin{tikzpicture}
                \def\r{1}
                \def\d{20}
                \def\dx{2.1}
                \def\hd{.5}

                \draw[->] (-\dx,0)++(90-\d:\r) arc (90-\d:-30+\d:\r);
                \draw[->] (-\dx,0)++(-30-\d:\r) arc (-30-\d:-150+\d:\r);
                \draw[->] (-\dx,0)++(-150-\d:\r) arc (-150-\d:-270+\d:\r);
                \draw (-\dx,0)++(90:\r) node (A) {$A$};
                \draw (-\dx,0)++(-30:\r) node (B) {$B$};
                \draw (-\dx,0)++(-150:\r) node (C) {$C$};

                \draw[->] (\dx,0)++(90-\d:\r) arc (90-\d:-30+\d:\r);
                \draw[->] (\dx,0)++(-30-\d:\r) arc (-30-\d:-150+\d:\r);
                \draw[->] (\dx,0)++(-150-\d:\r) arc (-150-\d:-270+\d:\r);
                \draw (\dx,0)++(90:\r) node (A2) {$A$};
                \draw (\dx,0)++(-30:\r) node (B2) {$B$};
                \draw (\dx,0)++(-150:\r) node (C2) {$C$};
                \draw (\dx,0)++(90-60:\r+.3) node (n1) {$\oplus 1$};
                \draw (\dx,0)++(-30-60:\r+.3) node (n2) {$\oplus 1$};
                \draw (\dx,0)++(-150-60:\r+.3) node (n3) {$\oplus 1$};

                \draw (-\hd/2,0) node (PLUS) {\LARGE $+$};
                \draw (-\dx-\r-\hd,0) node (half1) {\LARGE $\frac{1}{2}$};
                \draw (\dx-\r-\hd,0) node (half2) {\LARGE $\frac{1}{2}$};
        \end{tikzpicture}
        \caption{The circular identity channel  uniformly mixed with the circular bit-flip channel.}
        \label{fig:prob}
\end{figure}

In the probability view, there exists an example of a {\em randomized\/} process that results in non-causal correlations~\cite{aemin},  shown in Figure~\ref{fig:prob}.
In every run, the process models with probability~$1/2$ the clockwise identity channel or the clockwise bit-flip channel. 
In the probabilistic view, {\em both\/} channels appear with equal probability.
This leads to
every party's inability to influence its past.
For instance, if parties~$B$ and~$C$ copy the input to the output and party~$A$ has~$a$ on the output, then~$A$ has a random bit on the input~--- party~$A$ cannot influence its past, and the grandfather antinomy does not arise.
If, however, the probabilities of the mixture are altered slightly, a contradiction arises~\cite{njp}.
Furthermore, the process from Figure~\ref{fig:prob} cannot be embedded into a process with more inputs and outputs such that the larger process becomes {\em deterministic\/} and remains {\em logically consistent\/}~\cite{purification}.
Since in the  view studied here, we look at single runs {\em without\/} probabilities~--- either of the global relations from the left or from the right channel must hold.
Thus, if all parties use the universal local operation, a contradiction always arises, showing the inconsistency of the  process.

\

\noindent
{\it Discussion.}
In Section~\ref{eins} we saw that one consequence of dropping the notion of an {\em a priori\/} causal structure is that randomness becomes hard to define.
Thus, we are forced to take the ``{\em factual-only view\/}:'' No probabilities are involved.
Here, we show two facts that we formulate without considering {\em counterfactuals}.
The first is that causal relations among parties can be {\em derived\/} by considering fixed bit strings only, without the use of the probability language.
These causal relations are an {\em inherent\/} property of the bit strings of the parties.
In other words, these  strings are understood to be {\em logically
  prior\/} to the causal relations (just as in Section~\ref{spacom}).
The second consequence is that the causal relations that stem from certain strings can describe {\em non-causal\/} scenarios.
This means that {\em logical consistency\/} does not imply a  causal
scenario: 
{\em Causality is strictly stronger than logical consistency.}

\section{Conclusions}

Whereas for {\em Parmenides of Elea}, time was a mere illusion~| ``No
was nor will, all past and future null''~|, {\em Heraclitus\/} saw
space-time as the pre-set stage on which his play of
permanent change  starts and ends.  
The follow-up debate~| two millennia later and three centuries ago~| 
between {\em Newton\/} and {\em Leibniz\/}
about as how fundamental space and time, hence, {\em causality}, are to be
seen was decided by the course of science in favor of Newton: In
this 
view, space and time can be imagined as
fundamental and given {\em a priori}. (This applies also to
relativity theory, where space and time  get intertwined and dynamic
but  remain fundamental  instead of  becoming purely
relational in the sense of 
{\em Mach's principle}.) 
Today, we have more reason to question a fundamental causal structure~|
such 
as the difficulty of explaining quantum non-local correlations according 
to Reichenbach's principle. So motivated, we 
care to test
 refraining from assuming space-time as initially given; this has a number of consequences and
implications, some of which we address in this text. 

When causality is dropped, the usual definitions of randomness stop
 making sense. Motivated by this,
we test the use of intrinsic, context-independent ``randomness'' 
measures  such as a string's length minus its (normalized)
fuel value. 
We show that under the Church-Turing hypothesis, Kolmogorov 
complexity relates to this value. We argue that with respect to quantum
non-locality, complexity allows for a reasoning that avoids comparing 
results of different measurements that cannot all be actually carried
out, {\em i.e.}, that is {\em not counterfactual}. Some may see this as a 
conceptual simplification. It also leads to an all-or-nothing flavor 
of the Church-Turing hypothesis: {\em Either no physical system can
generate uncomputable sequences, or even a single photon can}.
Finally, it is asked whether {\em logical
reversibility\/} is connected to the second law of thermodynamics~|
interpreted here in complexities and independent of any context expressed through
probabilities
or ensembles~| 
and potentially to the arrow of time, past and future.
Finally, we have speculated that if
a causal structure is not
fundamental, how it may emerge from  data-compressibility relations.

When causality is dropped, one risks antinomies.
We show, in the complexity-based view,  that sticking to logical consistency 
does not restore causality but is strictly weaker. This observation
has recently been extended to {\em computational complexity\/}~\cite{cotton16,cotton162}:
Circuits solely avoiding antinomies are strictly stronger than causal circuits.

\begin{acknowledgments}
This text is based on a presentation at the ``Workshop on Time in
Physics,''
organized by {\em Sandra Rankovi{\'c}, Daniela Frauchiger, and  Renato Renner at
ETH Zurich\/} in Summer 2015. 

The authors thank 
Mateus Ara\'ujo, Veronika Baumann, Charles B\'edard, Gilles Brassard,
Harvey Brown, Caslav Brukner, Harry Buhrman, Matthias Christandl, Sandro Coretti, 
Fabio Costa, Bora Dakic, Fr\'ed\'eric Dupuis, 
Paul Erker, Adrien Feix, J\"urg Fr\"ohlich, Nicolas Gisin, 
Esther H\"anggi, Arne Hansen, Marcus Huber, Lorenzo Maccone, 
Alberto Montina, Samuel Ranellucci, Paul Raymond-Robichaud, Louis
Salvail, L.~Benno Salwey,  Andreas Winter, and Magdalena Zych
for inspiring discussions, and the Einstein Kaffee as well as the Reitschule Bern for their inspiring atmosphere.
{\em | Grazie mille!}

The authors thank {\em Claude Cr\'epeau\/} for his kind invitation to present
this work, among others, at the {\em 2016 Bellairs Workshop},  McGill
Research Centre, Barbados. 

	Our work was supported by the Swiss National Science
        Foundation (SNF), the National Centre of Competence in
        Research ``Quantum Science and Technology'' (QSIT), 
the COST action on Fundamental Problems in Quantum Physics, and the
Hasler Foundation.
\end{acknowledgments}


\begin{thebibliography}{10}


\bibitem{aare}
S.\ Aaronson,
http://www.scottaaronson.com/blog/?p= 762, 2012.

\bibitem{ara}
P.~K.~Aravind, 
Bell's theorem without inequalities and only two distant observers,
{\em  Foundations of Physics Letters}, Vol.\ 15, No.\ 4, pp.\ 397-–405, 2002.

\bibitem{jdb}
J.-D.\ Bancal, S.\ Pironio, A.\ Ac\'in, Y.-C.\ Liang, V.\ Scarani, N.\ Gisin,
Quantum non-locality based on finite-speed causal influences leads to superluminal signalling,
{\em Nature Physics}, Vol.\ 8, pp.\ 867--870, 2012.

\bibitem{tomy}
T.\ J.\ Barnea, J.-D.\ Bancal, Y.-C.\ Liang, N.\ Gisin,
Tripartite quantum state violating the hidden influence constraints,
{\em Physical Review A}, Vol.\ 88, pp.\ 022123, 2013.

\bibitem{kent}
J.\ Barrett, L.\ Hardy, A. Kent, 
No-signalling and quantum key distribution, 
{\em Physical Review Letters}, Vol.\ 95, pp.\ 010503, 2005.

\bibitem{aemin}
\"A.\ Baumeler, A.\ Feix, S.\ Wolf,
Maximal incompatibility of locally classical behavior and global causal order in multi-party scenarios,
{\em Physical Review A}, Vol.\ 90, pp.\ 042106, 2014.

\bibitem{isit}
\"A.\ Baumeler, S.\ Wolf,
Perfect signaling among three parties violating predefined causal order,
{\em Proceedings of IEEE International Symposium on Information Theory 2014}, pp.\ 526--530, IEEE, Piscataway, 2014.

\bibitem{purification}
\"A.\ Baumeler, F.\ Costa, T.\ C.\ Ralph, S.\ Wolf, M.\ Zych,
Reversible time travel with freedom of choice,
{\em arXiv preprint}, arXiv:1703.00779 [gr-qc], 2017.

\bibitem{njp}
\"A.\ Baumeler, S.\ Wolf,
The space of logically consistent classical processes without causal order,
{\em New Journal of Physics}, Vol.\ 18, pp.\ 013036, 2016.

\bibitem{cotton16}
\"A.\ Baumeler, S.\ Wolf,
Non-causal computation,
{\em Entropy}, Vol.\ 19, pp.~326, 2017.

\bibitem{cotton162}
\"A.\ Baumeler, S.\ Wolf,
Computational tameness of classical non-causal models,
{\em Proceedings of the Royal Society A}, Vol.\ 474, pp.\ 20170698, 2018.

\bibitem{bell}
J.\ S.\ Bell,
On the Einstein-Podolsky-Rosen paradox,
{\em Physics}, Vol.~1, pp.~195--200, 1964.

\bibitem{bennetttoc}
C.\ H.\ Bennett,
The thermodynamics of computation,
{\em International Journal of Theoretical Physics}, Vol.\ 21, No.\ 12, pp.\ 905--940, 1982.

\bibitem{bennettTM}
C.\ H.\ Bennett,
Logical reversibility of computation,
{\em IBM Journal of Research and Development}, Vol.\ 17, No.\ 6, pp.\ 525--532.

\bibitem{bbt}
G.\ Brassard, A.\ Broadbent, A.\ Tapp,
Quantum pseudo-telepathy, 
{\em Foundations of Physics}, Vol.\ 35, pp.~1877, 2005.

\bibitem{chaitin}
G.\ Chaitin,
A theory of program size formally identical to information theory,
{\em Journal of the ACM}, Vol.\ 22, pp.\ 329-–340, 1975.

\bibitem{cbc}
R.\ Cilibrasi, P.\ Vit\'anyi,
Clustering by compression,
{\em IEEE Transactions on Information Theory} Vol.~51, No.~4, pp.\ 1523-–1545, 2005. 

\bibitem{colren}
R.\ Colbeck, R.\ Renner,
No extension of quantum theory can have improved predictive power,
{\em Nature Communications}, Vol.\ 2, pp.\ 411, 2011.

\bibitem{colrenamp}
R.\ Colbeck, R.\ Renner,
Free randomness can be amplified,
{\em Nature Physics}, Vol.\ 8, pp.\ 450--454, 2012.

\bibitem{coretti}
S.\ Coretti, E.\ H\"anggi, S.\ Wolf,
Nonlocality is transitive,
{\em Physical Review Letters}, Vol.\ 107, pp.\ 100402, 2011.

\bibitem{dahlsten}
O.\ Dahlsten, R.\ Renner, E.\ Rieper, V.\ Vedral,
The work value of information,
{\em New Journal of Physics}, Vol.\ 13, pp.\ 053015, 2011.

\bibitem{everett}
H.\ Everett, ``Relative state'' formulation of quantum mechanics,
{\em Reviews of Modern Physics}, Vol.\ 29, No.\ 3, pp.\ 454--462, 1957.

\bibitem{fine}
A.\ Fine,
Hidden variables, joint probability, and the Bell inequalities,
{\em Physical Review Letters}, Vol.\ 48, pp.\ 291--295, 1982.

\bibitem{ballist}
E.\ Fredkin, T.\ Toffoli,
Conservative logic,
{\em International Journal of Theoretical Physics}, Vol.\ 21, No.\ 3--4, pp.\ 219-–253, 1982.

\bibitem{vitstat}
P.\ G\`{a}cs, J.\ T.\ Tromp, P.\ M.\ B.\ Vit\'{a}nyi,
Algorithmic statistics,
{\em IEEE Transactions on Information Theory}, Vol.\ 47, No.\ 6, pp.\ 2443--2463, 2001.

\bibitem{gisin}
N.\ Gisin, 
Time really passes, science can't deny that,
in {\em Time in Physics}, Birkh\"auser, Cham, 2017.

\bibitem{esth}
E.\ H\"anggi, R.\ Renner, S.\ Wolf,
Efficient information-theoretic secrecy from relativity theory,
{\em Proceedings of EUROCRYPT 2010}, LNCS, Springer-Verlag, 2010.

\bibitem{gh}
G.\ Hermann,
Die naturphilosophischen Grundlagen der Quantenmechanik,
{\em Abhandlungen der Fries'schen Schule}, Band 6, pp.\ 69--152, 1935.

\bibitem{CT}
S.\ C.\ Kleene, {\em Introduction to Metamathematics}, North-Holland, 1952. 

\bibitem{kol}
A.\ N.\ Kolmogorov,
Three approaches to the quantitative definition of information,
{\em Problemy Peredachi Informatsii}, Vol.~1, No.~1, pp.~3--11, 1965.

\bibitem{landau98}
R.\ Landauer,
Information is inevitably physical,
{\em Feynman and Computation 2}, 1998.
 
\bibitem{text}
M.\ Li, P.\ Vit\'{a}nyi,
{\em An Introduction to Kolmogorov Complexity and Its Applications}, Springer-Verlag, 2008. 

\bibitem{ocb}
O.\ Oreshkov, F.\ Costa, C.\ Brukner, 
Quantum correlations with no causal order,
{\em Nature Communications}, Vol.\ 3, pp.\ 1092, 2012.

\bibitem{christina}
O.\ Oreshkov, C.\ Giarmatzi,
Causal and causally separable processes,
{\em New Journal of Physics}, Vol.~18, pp.~093020, 2016.

\bibitem{pr}
S.\ Popescu, D.\ Rohrlich,
Quantum non-locality as an axiom,
{\em Foundations of Physics}, Vol.~24, pp.~379--385, 1994.

\bibitem{raz}
R.\ Raz,
A parallel repetition theorem,
{\em SIAM Journal on Computing}, Vol.~27, No.~3, pp.~763-–803, 1998.

\bibitem{Reichenbach}
H.\ Reichenbach,
The principle of the common cause,
in {\em The Direction of Time}, Chap.\ 19, pp.\ 157-–167, California Press, Berkeley, 1956.

\bibitem{renner}
R.\ Renner,
personal communication,
2013.

\bibitem{russell}
B.\ Russell,
On the notion of cause,
in {\em Proceedings of the Aristotelian Society}, New Series, Vol.~13, pp.~1--26, 1912.

\bibitem{specker}
E.\ Specker,
Die Logik nicht gleichzeitig entscheidbarer Aussagen,
{\em Dialectica}, Vol.\ 14, pp.\ 239--246, 1960. 

\bibitem{beforebefore}
A.\ Stefanov, H.\ Zbinden, N.\ Gisin, A.\ Suarez,
Quantum correlations with spacelike separated beam splitters in motion: Experimental test of multisimultaneity,
{\em Physical Review Letters}, Vol.\ 88, pp.\ 120404, 2002.

\bibitem{tit}
T.\ E.\ Stuart, J.\ A.\ Slater, R.\ Colbeck, R.\ Renner, W.\ Tittel,
An experimental test of all theories with predictive power beyond quantum theory,
{\em Physical Review Letters}, Vol.\ 109, pp.\ 020402, 2012.

\bibitem{szilard29}
L. Szil\'{a}rd,
\"Uber die Entropieverminderung in einem thermodynamischen System bei Eingriffen intelligenter Wesen (On the reduction of entropy in a thermodynamic system by the intervention of intelligent beings),
{\em  Zeitschrift f\"ur Physik}, Vol.\ 53, pp.\ 840--856, 1929.

\bibitem{wheeler}
J.\ A.\ Wheeler, Information, physics, quantum: the search for link,
{\em Proceedings III International Symposium on Foundations of Quantum
  Mechanics}, pp.~354--368, 1989.


\bibitem{witt}
L.\ Wittgenstein, 
Logisch-philosophische Abhandlung, {\em Annalen der Naturphilosophie},
Vol.~14, 1921.


\bibitem{pra}
S.\ Wolf,
Non-locality without counterfactual reasoning,
{\em Physical Review A}, Vol.\ 92, No.\ 5, pp.\ 052102, 2015.

\bibitem{Woodward}
J.\ Woodward,
{\em Making things happen: A theory of causal explanation},
Oxford University Press, Oxford, 2003.

\bibitem{ws}
C.\ Wood, R.\ Spekkens, 
The lesson of causal discovery algorithms for quantum correlations:
Causal explanations of Bell-inequality violations require fine-tuning,
{\em New Journal of Physics}, Vol.~17, pp.~033002, 2015.

\bibitem{zl}
J.\ Ziv, A.\ Lempel, Compression of individual sequences via variable-rate coding,
{\em  IEEE Transactions on Information Theory}, Vol.\ 24, No.\ 5, pp.\ 530--536, 1978.

\bibitem{bruz}
M.\ Zukowski, C.\ Brukner,
Quantum non-locality - It ain't necessarily so...,
{\em Journal of Physics A: Mathematical and Theoretical}, Vol.~47, pp.\ 424009, 2014.

\bibitem{zurek}
W.\ H.\ Zurek, 
Algorithmic randomness and physical entropy,
{\em Physical Review A}, Vol.\ 40, No.\ 8, pp.\ 4731--4751, 1989. 


\end{thebibliography}
\end{document}